\documentclass[12pt]{iopart}

\usepackage{iopams}  
\usepackage{graphicx}
\begin{document}

\title[Bouncing and Cyclic Quantum Primordial Universes and the Ordering Problem]{Bouncing and Cyclic Quantum Primordial Universes and the Ordering Problem}

\author{Isaac Torres$^{1}$, J\'ulio C. Fabris$^{2,3}$, Oliver F. Piattella$^{2,4}$}

\address{$^{1}$ PPGFis, CCE - Universidade Federal do Esp\'{i}rito Santo, 29075-910, Vit\'{o}ria, ES, Brazil}
\address{$^{2}$ N\'{u}cleo Cosmo-UFES \& Departamento de F\'{i}sica - Universidade Federal do Esp\'{i}rito Santo 29075-910, Vit\'{o}ria, ES, Brazil}
\address{$^{3}$ National Research Nuclear University MEPhI, Kashirskoe sh. 31, Moscow 115409, Russia}
\address{$^{4}$ Institut f\"ur Theoretische Physik, Ruprecht-Karls-Universit\"at Heidelberg, Philosophenweg 16, 69120 Heidelberg, Germany}

\ead{itsufpa@gmail.com}
\vspace{10pt}
\begin{indented}
\item[]November 2019
\end{indented}

\begin{abstract}
In a Bohmian quantum cosmology scenario, we investigate some quantum effects on the evolution of the primordial universe arising from the adoption of an alternative non-trivial ordering to the quantization of the constrained Hamiltonian of a minimally coupled scalar field. The Wheeler-DeWitt equation has a contribution from the change in factor ordering, hence there are new quantum effects. We compare the results between the non-trivial and the trivial ordering cases, showing that the classical limit is valid for both orderings, but new bouncing and cyclic solutions are present in the non-trivial case. Additionally, we show that the non-singular solutions already present in the trivial ordering formalism keep valid. 
\end{abstract}


%
%

\section{Introduction}
The recent discovery of the accelerated expansion of the universe \cite{Riess:1998cb, Perlmutter:1998np} has made gravitational theories based on a minimally-coupled scalar field one of the most well-studied class of models in cosmology \cite{Ratra:1987rm, Caldwell:1997ii}. For a canonical scalar field $\phi$, the Lagrangian has the following form \cite{Faraoni:2004pi}:
\begin{equation}\label{l}
	L = \sqrt{-g}\left[R -\frac{1}{2}g^{\mu\nu}\frac{\partial\phi}{\partial x^\mu}\frac{\partial\phi}{\partial x^\nu} - V(\phi)\right]\;,
\end{equation}
where $R$ is the Ricci scalar and $V(\phi)$ is some potential. It is specially important for the cosmology of the very early universe \cite{Linde1990xn}, when quantum effects become fundamental. Those quantum effects may be a solution for the singularity problem.

There are, in fact, several different theories and interpretations of the primordial quantum universe, and one of them is Bohm-de Broglie (also known as Bohmian) quantum cosmology. In \cite{PhysRevD.57.4707, PINTONETO2000194, Colistete:2000ix}, it is described how such an alternative interpretation can be generalized from quantum mechanics to cosmological models such as those of \eref{l}. In particular, even when $V = 0$ one is able to solve the singularity problem thanks to quantum corrections, which induce a bounce. The case of an exponential potential, related with a matter-dominated universe, was recently presented in \cite{PhysRevD.96.063502}.

The Lagrangian  \eref{l} can also be seen as one of the simplest particular cases of Horndeski modified gravity theory \cite{Horndeski:1974wa}, thus it is free of Ostrogradsky instability \cite{Ostrogradsky:1850fid, Kobayashi_2019}. It is also in agreement with the recent constraints imposed by GW170817 and GRB170817A \cite{PhysRevLett.119.161101, Goldstein_2017, Abbott_2017} on the velocity of gravitational waves \cite{Kobayashi_2019}. Lagrangian  \eref{l} is also related with effective string theory \cite{PhysRevD.57.4707} and is also the Einstein frame version of several other scalar-tensor theories of gravity \cite{Faraoni:2004pi}.

The motivation for adopting an alternative interpretation of quantum mechanics in a cosmological setting comes from the fact that the exterior domain hypothesis \cite{Omnes}, tacitly present in most standard interpretations of quantum mechanics, is considered by some authors as being a conceptual problem when the system under investigation is the universe \cite{ACACIODEBARROS1998229, 10.2307/193027, PintoNeto:2004uf}. Because of that, it has been proposed \cite{Blaut_1996,Barros_1997,PintoNeto:2004uf, PINTO-NETO2000} to adopt in cosmology the Bohmian interpretation of quantum mechanics \cite{PhysRev.85.166, PhysRev.85.180}. Besides that, the Bohmian interpretation also solves the problem of time ambiguity in quantum cosmology and quantum gravity \cite{Kiefer:2013jqa, andprobtime}, thanks to the guidance equations.  More about Bohm-de Broglie quantum mechanics can be found in \cite{cushing2013bohmian, durr2009bohmian, holland_1993, freire2014quantum, pladevall2019applied, HOLLAND199395}.  

Another conceptual problem faced in the quantization of a gravitational theory like \eref{l} is the factor ordering ambiguity, a direct consequence of Dirac's quantization rule. In \cite{PINTO-NETO2000}, it is shown that the basic features of Bohmian quantum gravity do not really depend on the factor ordering, although some ordering must be chosen to actually apply quantization. Basead on that argument, it is common to apply only the trivial ordering in the quantization of the constrained Hamiltonian, like it is done in \cite{PINTO-NETO2000}. But the factor ordering ambiguity remains, so that we can ask ourselves: if the general features of that theory are invariant under a change of ordering, what happens with the time evolution of the scale factor for a non-trivial ordering? 

An interesting non-trivial ordering was proposed by T. Christodoulakis and J. Zanelli in \cite{PhysRevD.29.2738, Christodoulakis1986}. Their main idea is to avoid the ordering ambiguity by introducing a canonical transformation that makes the Hamiltonian assume a very simple form in which there is no ambiguity. Thus, returning to the old variables, a natural factor ordering arises. In standard interpretation of quantum mechanics, that non-trivial ordering implies that the Hamiltonian is quantized as a Hermitian operator, but in Bohmian quantum cosmology it is not clear what would be the consequences of assuming such an ordering. In this paper, we propose a generalization of that non-trivial ordering to $N$ dimensions (see  \eref{ordit}). In fact, it was applied to a model similar to \eref{l} in \cite{Maeda_2015}, with a standard quantum interpretation, giving a non-singular expected value for the scale factor $a$.

As mentioned above, Bohmian quantum cosmology has been applied to Lagrangian  \eref{l} quite successfully, in the sense that both the problem of time and the singularity problem can be avoided. The former is solved by the guidance equations and the latter by the choice of a non-trivial complex wave solution to Wheeler-DeWitt equation. Thus, it is important to understand complementary aspects of that theory, such as how another conceptual problem of quantum cosmology, namely, the ordering ambiguity, related with quantization itself, is dealt in Bohmian quantum theories. Since our goal in this paper is to explore the ordering ambiguity, we do not enter the debate on the interpretation of the quantum theory, which is still open. For more about the latter we refer the reader to e.g. \cite{freire2014quantum}. In this paper, we study in detail what are the consequences of the non-trivial ordering  \eref{ordit} for the Bohmian quantum cosmology of the Lagrangian  \eref{l}, assuming $V = 0$ and focusing only on the minisuperspace of the Friedmann-Lema\^itre-Robertson-Walker metrics and how quantum effects modify the evolution of the scale factor.

In section \ref{revw}, we review the Bohmian formalism of \cite{PhysRevD.57.4707, PINTONETO2000194, Colistete:2000ix} in order to give the basis for properly comparing the differences between the formalisms for the two orderings. In section \ref{ordprob}, we describe the ordering problem in quantum cosmology and present the solution proposed in \cite{PhysRevD.29.2738, Christodoulakis1986} and its generalization  \eref{ordit}. This leads to the modified Wheeler-DeWitt equation  \eref{eqwdwz}, presented in section \ref{modwdw}. The link between the two orderings becomes clearer with the introduction of an ordering parameter $r \geq 0$ that allows a continuous transition between them. We also show in section \ref{modwdw} that the connection between classical and quantum dynamics is independent from $r$. In section \ref{new}, we show how the non-trivial ordering allows new bouncing and cyclic universe solutions, whose would degenerate to singular ones for the trivial ordering. In section \ref{old}, we study the modifications of the bouncing and cyclic solutions existing for the trivial ordering. The formalism for the trivial ordering already admits bouncing and cyclic solutions. In section \ref{old}, we study the effect of the ordering change over those solutions, showing that the non-trivial ordering leads to bouncing and cyclic solutions very similar to the old ones. Therefore, the non-trivial ordering proposed here maintains old non-singular solutions but also furnishes new ones. Finally, section \ref{conc} is devoted to our conclusions.


\section{Review of Bohmian quantum cosmology}\label{revw}
Let us briefly review the application of the Bohmian approach to quantum cosmology for the theory represented by the Lagrangian  \eref{l}, for a vanishing potential, studied in \cite{PhysRevD.57.4707, PINTONETO2000194, Colistete:2000ix}. After rescaling $\phi$, it is equivalent to:
\begin{equation}\label{ham}
H=N\mathcal{H}=\frac{\kappa^2N}{12V\rm e^{3\alpha}}\left( -p_{\alpha}^{2}+p_{\phi }^{2}\right)\;,
\end{equation}
where $H$ is the (constrained) Hamiltonian, $V$ and $\kappa^2\equiv8\pi G$ are constants (V must not be confused with the potential in  \eref{l} and $G$ is Newton's gravitational constant), $\phi (t)$ is the minimally coupled scalar field, $\alpha(t)\equiv\ln a(t)$, where $a(t)$ is the scale factor, $p_{\alpha},p_{\phi}$ are the canonical conjugated momenta, and $N(t)$ is the lapse function \cite{Calcagni:2017sdq}, for a FLRW background:
\begin{equation}\label{flrw}
ds^2 = -N(t)^2dt^2 + a(t)^2\delta_{ij}dx^idx^j\;,
\end{equation}
where we can see that the curvature of the spatial sections is flat. The constant $V$ represents the volume of the conformal hypersurface, which is choosen to be $V\equiv4\pi l_{\rm P}^{3}/3$, where $l_{\rm P}$ is the Planck length, in order to guarantee that when $a=1$ (which is equivalent to $\alpha=0$), the universe has approximately the Planck volume, as mentioned in \cite{PhysRevD.97.083517}. This is how we can guarantee that the analysis developed here takes place in a quantum scale. As we shall see, this argument is valid for both the trivial and the non-trivial orderings. By the values of $V$ and $\kappa$ above, we can see that the global constant in $H$ is $G/2l_{\rm P}^{3}$. Now, in units such that $c=\hbar=1$, which we shall adopt here, $l_{\rm P}^{2}=G$, so that the global factor in \eref{ham} is:
\begin{equation}\label{lp}
	\frac{\kappa^2}{12V}=\frac{1}{2l_{\rm P}}\;.
\end{equation}

 Representing the time derivative with a dot, the Hamilton equations for $\dot{\alpha}$ and $\dot{\phi}$ are the following:
\begin{equation}\label{momc}
\eqalign{\dot{\alpha}=-\frac{N}{l_{\rm P}}{\rm e}^{-3\alpha}p_{\alpha}\;, \cr
	\dot{\phi }=\frac{N}{l_{\rm P}}{\rm e}^{-3\alpha}p_{\phi }\;,}
\end{equation}
whilst the Hamilton equation for $N$ gives the constraint $\dot{\alpha}^2 = \dot{\phi }^2$. After deriving the constraint,  we can set $N=1$ and easily obtain the classical dynamics of $\alpha$ and $\phi$:
\begin{equation}\label{dinclass}
\eqalign{\ddot{\alpha}+3\dot{\alpha}^2 = 0\;,\cr
	\ddot{\phi }+3\dot{\alpha}\dot{\phi } = 0\;.}
\end{equation}
It thus follows that the classical scale factor time evolution is singular: $a(t)=(t/t_0)^{1/3}$, where $t_0$ is the age of the universe. Therefore, in the theory represented by \eref{ham}, in order to obtain a non-singular solution, we must look for a quantum correction. 

In \cite{PINTONETO2000194, Colistete:2000ix}, Hamiltonian \eref{ham} is quantized by applying the usual Dirac rule with the trivial ordering
\begin{equation}\label{do}
\frac{1}{2}f(q)p^{2}\longrightarrow \frac{1}{2}f(q)\hat{p}^{2}=-\frac{\hbar^2}{2}f(q)\frac{\partial^2}{\partial q^{2}}\;,
\end{equation}
thus leading to the Wheeler-DeWitt equation, $\hat{\mathcal{H}}\psi = 0$:
\begin{equation}\label{wdwdo}
\frac{\partial^2\psi}{\partial\alpha^2} -\frac{\partial^2\psi}{\partial\phi ^2}=0\;,
\end{equation}
where $\psi = \psi(\alpha,\phi )$ represents the so-called wave function of the universe. In \eref{do}, $q$ and $p$ are the generalized coordinates and momenta, respectively. We refer to  \eref{do} as the trivial ordering because it is the simplest choice: to arrange scalars as just the coefficients of the differential operators. The general solution of \eref{wdwdo} is the D'Alembert solution:
\begin{equation}\label{wdwdogs}
\psi(\alpha,\phi ) = F(\phi +\alpha)+G(\phi -\alpha)\;,
\end{equation}
where $F$ and $G$ are generic $C^1$ functions. The simplest complex wave solution of \eref{wdwdo} is:
\begin{equation}\label{wdwpw}
\psi_k(\alpha,\phi) = {\rm e}^{ik(\phi \pm\alpha)}\;,
\end{equation}
where $k$ is a real separation constant. As we explain at the end of this section, \eref{wdwpw} can be considered a trivial solution in the Bohmian interpretation. An example of non-trivial solution is the Gaussian wave packet:
\begin{equation}\label{wdwpo}
\psi(\alpha,\phi )=\int {\rm d}k\: {\rm e}^{-\frac{(k-k_0)^2}{\sigma^2}}\left[ {\rm e}^{ik(\phi -\alpha)}+{\rm e}^{ik(\phi +\alpha)}\right]\;,
\end{equation}
where $k_0$ and $\sigma$ are constants.

In order to interpret a given wave solution $\psi$ using Bohmian formalism, we must express it in polar form: $\psi=R\exp(\rm i S/\hbar)$, where $R$ and $S$ are real functions. Thus, the real part of equation \eref{wdwdo} becomes
\begin{equation}\label{wdwdore}
\frac{\kappa^2{\rm e}^{-3\alpha}}{12V}\Bigg[ -\left(\frac{\partial S}{\partial\alpha}\right)^{2} + \left(\frac{\partial S}{\partial\phi }\right)^{2} \Bigg] + Q(\alpha,\phi )=0\;,
\end{equation}
where 
\begin{equation}\label{wdwdopq}
Q(\alpha,\phi )=\frac{\kappa^2\hbar^2}{12V}\frac{{\rm e}^{-3\alpha}}{R}\left(\frac{\partial^2R}{\partial\alpha^2} -\frac{\partial^2R}{\partial\phi ^2}\right)\;.
\end{equation}
Thus, \eref{wdwdore} is the Hamilton-Jacobi equation associated to \eref{ham}, except for the term $Q(\alpha,\phi )$. In a Bohmian perspective, this means that the phase function $S$ plays the role of the Hamilton principal function, from which follow the relations:
\begin{equation}\label{guid}
p_{\alpha}=\frac{\partial S}{\partial\alpha}\;, \qquad p_{\phi }=\frac{\partial S}{\partial\phi }\;,
\end{equation}
called the ``guidance equations''. Comparing now \eref{momc} and \eref{guid}, we see that the guidance equations are invariant under a time reparametrization, which means that Bohmian quantum cosmology solves the problem of time, as it was shown in \cite{doi:10.1142/S0218271898000164}. It also follows that the additional term $Q$ is a quantum contribution (of order $\hbar^2$) to the classical Hamilton-Jacobi equation associated with \eref{ham}. Thus, $Q$ is called the ``quantum potential'' associated with \eref{ham}, in analogy with the Hamilton-Jacobi theory and Bohmian quantum mechanics \cite{holland_1993}. With that interpretation, the wave function is also called the ``pilot wave'' that guides the solutions for $\alpha$ and $\phi$. The guidance equations can be rewritten using \eref{momc} to become an autonomous dynamical system for $\alpha$ and $\phi $, taking $N = 1$ (i.e. using the cosmic time):
\begin{equation}\label{distdinonger}
		\eqalign{l_P\dot{\alpha} =-{\rm e}^{-3\alpha}\frac{\partial S}{\partial\alpha}\;,\cr
		l_P\dot{\phi } ={\rm e}^{-3\alpha}\frac{\partial S}{\partial\phi }\;.}
\end{equation}
We must stress that \eref{distdinonger} is indeed a quantum system, because it clearly dominates at the Planck scale, but also because the classical Hamilton principal function is replaced by the phase $S$ of a wave function. As we comment below, they are equal only in the particular case for which the quantum potential vanishes. When quantum effects occur, they must be different.

Although $\psi$ is a combination of $R$ and $S$, it is easier to solve the Wheeler-DeWitt equation than to solve for $R$ and $S$ directly. After finding the pilot wave $\psi$, it becomes straightforward to find $R$ and $\partial S/\partial q$, because $\psi=R\exp(\rm i S/\hbar)$ implies
\begin{equation}\label{r-e-s}
R=\sqrt{\psi^*\psi}\;, \qquad \mbox{and} \qquad\frac{\partial S}{\partial q} =\hbar\:{\rm Im}\left(\frac{1}{\psi}\frac{\partial\psi}{\partial q}\right)\;.
\end{equation}
Hence, given a wave function, we use \eref{r-e-s} in order to determine both the quantum potential and the guidance equations. With this formalism, one can now study the quantum dynamics of a given wave function.

Putting the plane wave solution  \eref{wdwpw} into \eref{distdinonger} and taking the time derivative of the result, we can see that the classical equations of motion \eref{dinclass} are recovered. Thus, the plane wave solution gives no quantum contribution. That result agrees also with the meaning of $Q$: from \eref{wdwdopq}, we see that a plane wave gives a null quantum potential. Therefore, a nontrivial quantum contribution requires a more complicated solution. For example, for the wave packet  \eref{wdwpo}, there is a non-trivial quantum potential, thus inducing a deviation of \eref{distdinonger} from the classical case. As shown in \cite{PINTONETO2000194, Colistete:2000ix}, this correction generate bouncing universes, an important class of non-singular cosmological solutions for $a(t)$ \cite{NOVELLO2008127}. Since $Q \sim \hbar^2a^{-3}$, this is a pure quantum effect, so that the initial singularity is avoided and the classical evolution  \eref{dinclass} is asymptotically recovered.

The quantum theory reviewed in this section was further developed in order to describe, for example, creation of particles \cite{PhysRevD.95.023523}, cosmological perturbations \cite{PhysRevD.97.083517}, and primordial gravitational waves \cite{PhysRevD.95.123522}. We do not discuss here those results, limiting our analysis to the ordering ambiguity and its consequences for the above aspects of the theory.


\section{The ordering problem}\label{ordprob}
Dirac's quantization rule prescribes that both generalized  coordinates $q_m$ and momenta $p_n$ in the Hamiltonian must be replaced by linear operators acting on the wave function $\psi$. Those operators are defined by:
\begin{equation}\label{dirac}
\hat{q}_m\psi = q_m\psi\;, \qquad \hat{p}_n\psi=-{\rm i}\hbar\partial_n\psi\;,
\end{equation}
from which the usual commutation relation follows:
\begin{equation}\label{commrel}
[\hat{q}_m,\hat{p}_n]={\rm i}\hbar\delta_{mn}\;.
\end{equation}
Then, if the (one-dimensional, for simplicity) Hamiltonian $H(q,p)$ is as simple as 
\begin{equation}\label{hamgen}
H=\textstyle\frac{1}{2}f(q)p^2\;,
\end{equation}
it follows from the commutation relations above that Dirac's quantization rule \eref{dirac} is ambiguous, because we do not know what is the right ordering of $f(q)$ and $p^2$ for which we should apply \eref{dirac}. This is the ordering problem. If $f$ is constant, which is the case for systems of particles in basic quantum mechanics, then the ambiguity disappears. But that is not the general case for quantum cosmology, even for a minimal coupling like that in \eref{ham}, for which $f =\exp(-3\alpha)$. Therefore, any quantization of the Hamiltonian \eref{ham} implicitly assumes a particular choice of ordering and the number of possibilities are actually infinite. The most obvious choice is \eref{do}, the trivial ordering.

Several different criteria to solve the ordering ambiguity have been proposed in the literature, in the context of quantum cosmology and quantum gravity. For example, see \cite{PhysRev.114.1182, PhysRevD.20.830, PhysRevD.29.2738, Christodoulakis1986, PhysRevD.37.888, PhysRevA.41.1199, PhysRevD.59.063513, PhysRevD.80.103507, PhysRevD.100.046008,Steigl:2005fk}. Additionally, for complementary mathematical aspects of the ordering problem, see for instance \cite{gotay1999, Chithiika} and references therein. Each criterion has its specific physical (or mathematical) motivations, and gives rise to different quantum effects. Because of that, there is no definitive answer for what should be the right choice. 

One particularly interesting way to avoid ordering ambiguity, first proposed in \cite{PhysRevD.29.2738,Christodoulakis1986}, is to find the Lagrangian equivalent to \eref{hamgen}, and then define $q'= \int f^{-1/2}(q)\rmd q$, which implies, after some algebra, that $p'=f^{1/2}p$. Hence, we end up with the transformed Hamiltonian $H'=\textstyle\frac{1}{2}p'^{2}$, which has no ordering ambiguity. Even though a detailed proof can be found in \cite{Christodoulakis1986}, we can see the equivalence between $H$ and $H'$ also by means of the canonical transformation obtained from the generating function
\begin{equation}\label{funcger}
	F(q,P)=P\int \, f^{-1/2}(q)\rmd q\;,
\end{equation}
where $(q,p)$ are the old coordinate and momentum and $(Q,P)$ are the new ones. They are related by the rules \cite{goldstein2002classical}:
\begin{equation}
	p=\frac{\partial F}{\partial q}\;,\qquad Q=\frac{\partial F}{\partial P},\qquad\mbox{and}\qquad H'(Q,P)=H(q,p)\;.
\end{equation}

Now, the quantization of $H'$ leads unavoidably to the following rule, for the old variables, $q$ and $p$:
\begin{equation}\label{ordzan}
\frac{1}{2}f(q)p^{2}\longrightarrow \frac{1}{2}f^{1/2}(q)\hat{p}f^{1/2}(q)\hat{p}=-\frac{\hbar^2}{2}f^{1/2} \frac{\partial}{\partial q} \left(f^{1/2} \frac{\partial}{\partial q}\right)\; .
\end{equation}
To see why this rule follows from the quantization of $H'$, first note that \eref{dirac} implies that
\begin{equation}
	\hat{p}'\psi=-{\rm i}\hbar\frac{\partial\psi}{\partial q'}=-{\rm i}\hbar\frac{\partial\psi}{\partial q}\frac{{\rm d} q}{{\rm d} q'}\;.
\end{equation}
Now, since ${\rm d}q'/{\rm d} q=f^{-1/2}(q)$, it follows from the rule for the derivative of the inverse function that ${\rm d}q/{\rm d} q'=f^{1/2}(q)$. Thus, 
\begin{equation}
	\hat{p}'\psi=f^{1/2}\left( -{\rm i}\hbar\frac{\partial\psi}{\partial q} \right)=f^{1/2}\hat{p}\,\psi\;.
\end{equation}
Therefore, taking $\Phi\equiv\hat{p}'\,\psi$, the rule above, when applied to $\Phi$, says that
\begin{equation}
	\hat{p}'^{2}\psi=\hat{p}'\Phi=f^{1/2}\hat{p}\,\Phi=f^{1/2}\hat{p}\,\hat{p}'\psi=f^{1/2}\,\hat{p}\,f^{1/2}\,\hat{p}\,\psi\;,
\end{equation}
thus proving \eref{ordzan}.

In \cite{Christodoulakis1986}, a generalization of  \eref{ordzan} to N dimensions is presented. In this paper, we propose an alternative generalization of \eref{ordzan}, for the case of interest. The easiest way to generalize the main idea of \eref{ordzan} is by taking now
\begin{equation}\label{funcger2}
F(q,P)=\frac{2}{3}e^{3\alpha/2}P_1+\phi P_2+NP_3\;
\end{equation}
as the generating function. Then, by the relations below \cite{goldstein2002classical},
\begin{equation}\label{cap3-eq-transf-can-ddim}
p_j=\frac{\partial F}{\partial q_j}\;,\qquad Q_j=\frac{\partial F}{\partial P_j},\qquad\mbox{and}\qquad H'(Q,P)=H(q,p)\;,
\end{equation}
we find the new Hamiltonian:
\begin{equation}\label{ham2}
H'(Q,P)=Q_3\left( -\frac{1}{2}P_{1}^{2}+\frac{2}{9Q_{1}^{2}}P_{2}^{2} \right)\;.
\end{equation}
Now, since the canonical transformations preserve the Hamilton equations of motion, $H'$ is entirely equivalent to $H$. Thus, by the canonical quantization rule when applied to \eref{ham2}, it follows that $[\hat{Q}_m,\hat{P}_n]={\rm i}\hbar\delta_{mn}$, which implies that there is no ordering ambiguity in the quantization of \eref{ham2}, since for all the products $\hat{Q}_m\hat{P}_n$ that appear in \eref{ham2} we have $m\neq n$. Moreover, the Hamiltonian constraint is also preserved, because \eref{ham2} can be written in the form $H'=Q_3\mathcal{H}'$, where $\mathcal{H}'$ does not depend on $Q_3=N$. Therefore, the Hamilton equations for $H'$ give the constraint $\mathcal{H}'\approx0$, thus leading to the Wheeler-DeWitt equation $\hat{\mathcal{H}}'\psi=0$, that is,
\begin{equation}\label{wdwzanvarmod}
	\frac{1}{2}\frac{\partial^2\psi}{\partial Q_{1}^{2}}- \frac{2}{9Q_{1}^{2}}\frac{\partial^2\psi}{\partial Q_{2}^{2}}=0\;.
\end{equation}

Hence, by a process entirely analogous to the one dimensional case, we conclude that the Wheeler-DeWitt equation \eref{wdwzanvarmod} is transformed into
\begin{equation}\label{wdwzanvarorig}
\frac{\partial^2\psi}{\partial\alpha^2} -\frac{3}{2}\frac{\partial\psi}{\partial\alpha} -\frac{\partial^2\psi}{\partial\phi ^2}=0\;.
\end{equation}
Therefore, this is entirely equivalent to consider the following ordering in the quantization of \eref{ham}:
\begin{equation}\label{ordit}
\hat{H}=-\frac{1}{2}f^{1/2}\hat{p}_{\alpha}f^{1/2}\hat{p}_{\alpha}+ \frac{1}{2}f^{1/2}\hat{p}_{\phi}f^{1/2}\hat{p}_{\phi} \; ,
\end{equation} 
where $f(\alpha)\equiv\exp(-3\alpha)$. This shows that \eref{ordit} is a natural generalization of \eref{ordzan} for the cosmological minisuperspace Hamiltonian \eref{ham}. This is the non-trivial ordering we will consider here. Observe also that applying \eref{ordit} is equivalent to just apply \eref{ordzan} for each coordinate, taking into account the sign of $\pm f$ in order to avoid imaginary variables. Finally, observe that, introducing an ordering parameter $r\geq0$, both the nontrivial ordering \eref{ordit} and the trivial one \eref{do} become particular cases of 
\begin{equation}\label{orditr}
\hat{H}=-\frac{1}{2}f^{1-r}\hat{p}_{\alpha}f^{r}\hat{p}_{\alpha}+ \frac{1}{2}f^{1-r}\hat{p}_{\phi}f^{r}\hat{p}_{\phi} \; ,
\end{equation}
where $r=0$ corresponds to the trivial ordering and $r=1/2$ corresponds to the nontrivial one, \eref{ordit}.

The ordering \eref{ordit} is equivalent, for some cases, to the Laplace-Beltrami ordering, as it is the case for \cite{Maeda_2015}, in which a cosmological theory similar to the one represented by \eref{ham} is considered, in the standard interpretation of quantum mechanics, and a non-singular expected value $\langle a^3\rangle$ is obtained. In the next sections, we apply ordering \eref{orditr} to Hamiltonian \eref{ham}, showing that the non-singular solutions already present in the $r=0$ case are maintained, in some sense, but there are also new bouncing and cyclic solutions for $r>0$, with a focus on $r=1/2$.


\section{Wheeler-DeWitt equation for the non-trivial ordering}\label{modwdw}
In order to compare the dynamics of the two orderings, we apply the ordering \eref{orditr} to Hamiltonian \eref{ham}, and hence the Wheeler-DeWitt equation is now, for any $r\geq0$:
\begin{equation}\label{eqwdwz}
\frac{\partial^2\psi}{\partial\alpha^2} -3r\frac{\partial\psi}{\partial\alpha} -\frac{\partial^2\psi}{\partial\phi ^2}=0\;.
\end{equation}
The change of ordering only acts over the $\alpha$-terms because the corresponding $f(q)$ of our problem does not depend on $\phi$, and since $\alpha$ and $p_{\phi}$ commute, we have $[f(\hat{q}),\hat{p}_{\phi}]=0$. Therefore, there is no possible ordering ambiguity in the quantization of the second term of \eref{ham}.

In comparison with \eref{wdwdo}, we see that the contribution of the change of ordering is the first-order derivative term. For any $r\neq0$, this term breaks the D'Alembert symmetric solution  \eref{wdwdogs}, thus changing the quantum evolution of $\alpha$ and $\phi $. But, before solving \eref{eqwdwz}, it is necessary to clarify how the Bohmian interpretation can be applied to this modified equation. Writing $\psi$ in the polar form $\psi=R\exp(\rm i S/\hbar)$, where $R(\alpha,\phi )$ and $S(\alpha,\phi )$ are real functions, the real part of \eref{eqwdwz} becomes equal to \eref{wdwdore} in form, but the quantum potential is now given by
\begin{equation}\label{pqz}
Q(\alpha,\phi )=\frac{\kappa^2\hbar^2}{12V}\cdot\frac{{\rm e}^{-3\alpha}}{R}\left(\frac{\partial^2R}{\partial\alpha^2}-3r \frac{\partial R}{\partial\alpha}- \frac{\partial^2R}{\partial\phi ^2}\right)\;.
\end{equation}
From the Hamilton-Jacobi structure of \eref{wdwdore}, we conclude that the guidance equations \eref{guid} (and therefore \eref{distdinonger} and \eref{r-e-s}) are also valid for \eref{orditr} and, in particular, for $r=1/2$.

On the other hand, from \eref{wdwdopq} and \eref{pqz} we can see that the quantum potential has a contribution due to the change of ordering. This is in accordance with what is expected from a quantum theory: since the ordering problem is an ambiguity in quantization, it is quite natural that any change of ordering affects the quantum dynamics only. We know that the intensity of the quantum potential tells us where in the phase space the quantum effects are more significant and where the classical dynamics is recovered. Thus, we can expect from the change in the quantum potential that new quantum effects arise. We shall see that this is indeed the case and we shall explore what are the implications for the quantum Bohmian trajectories and for the time evolution of the scale factor. From now on, we adopt units such that $\hbar=1$. The $\hbar$ was explicitly written until now just to evidence that the meaning of the quantum potential is still valid after the change of ordering.

Equation \eref{eqwdwz} can be separated by writing $\psi(\alpha,\phi )=A(\alpha)F(\phi )$, thus leading to:
\begin{equation}\label{eqsep}
\frac{A''(\alpha)}{A(\alpha)}-3r\frac{A'(\alpha)}{A(\alpha)}= \frac{F''(\phi )}{F(\phi )}\equiv\pm k^2\;,
\end{equation}
where $k$ is a real separation constant. The sign determines if the solutions are real or oscillatory complex waves. In what follows, we denote as $c_i$ ($i = 1,2,\dots$) the integration constants. The simplest possible case is $k=0$:
\begin{equation}\label{keq0}
\eqalign{A(\alpha)=c_1+c_2{\rm e}^{3r\alpha}\;,\cr
	F(\phi ) =c_3+c_4\phi\;.}
\end{equation}
If we choose $+k^2$, $k\neq0$, only real solutions are found:
\begin{equation}\label{mk2}
\eqalign{A_k(\alpha)={\rm e}^{3r\alpha/2}\left(c_1{\rm e}^{\sqrt{k^2+(3r/2)^2}\alpha} +c_2{\rm e}^{-\sqrt{k^2+(3r/2)^2}\alpha}\right)\;,\cr
	F_k(\phi )=c_3{\rm e}^{k\phi }+c_4{\rm e}^{-k\phi }\;. }
\end{equation}
Finally, choosing $-k^2$, $k\neq0$, we find:
\begin{equation}\label{mmk2}
\eqalign{A_k(\alpha)= {\rm e}^{3r\alpha/2}\left(c_1{\rm e}^{{\rm i}\omega\alpha} +c_2{\rm e}^{-{\rm i}\omega\alpha}\right)\;, \cr
	F_k(\phi )=c_3{\rm e}^{{\rm i}k\phi }+c_4{\rm e}^{-{\rm i}k\phi }\;, }
\end{equation}
where 
\begin{equation}\label{omega}
\omega\equiv\sqrt{k^2-(3r/2)^2}>0\;.
\end{equation}
In this case, in order to obtain oscillatory solutions in $\alpha$, $k$ must satisfy $|k|>3r/2$, which implies $\omega>0$. If $0<|k'|<3r/2$ (denoted $k'$ to avoid confusion), $A_{k'}(\alpha)$ is the real function
\begin{equation}\label{akl}
A_{k'}(\alpha)= {\rm e}^{3\alpha/4}(c_1{\rm e}^{\omega'\alpha}+c_2{\rm e}^{-\omega'\alpha})\;,
\end{equation}
where
\begin{equation}\label{omegal}
\omega'\equiv\sqrt{(3r/2)^2-k'^2}>0\;,
\end{equation}
with $F_{k'}$ given by \eref{mmk2}. Only real solutions are found for the particular cases $k=\pm3r/2$.

From the linearity of \eref{eqwdwz}, it follows that we can take linear combinations of the solutions above to construct other ones. Among all possibilities, we study the dynamics of five representative solutions.

\subsection{Recovering Classical Universe Dynamics}
The generalization of the plane wave solution \eref{wdwpw} for the modified Wheeler-DeWitt equation \eref{eqwdwz} is:
\begin{equation}\label{psis}
\psi_{\rm S}={\rm e}^{({\rm i}\omega+3r/2)\alpha+{\rm i}k\phi }\;,
\end{equation}
obtained from \eref{mmk2}. The subscript stands for ``singular''. It thus follows from \eref{guid},  \eref{distdinonger} and  \eref{r-e-s} that, for the single wave solution \eref{psis} (for any $r\geq0$), the quantum dynamical system is just:
\begin{equation}\label{sdqss}
\eqalign{\dot{\alpha}=-\omega {\rm e}^{-3\alpha}\;,\cr
	\dot{\phi }=k {\rm e}^{-3\alpha}\;.}
\end{equation}
Then, taking the time derivative of \eref{sdqss}, we see that the classical dynamics \eref{dinclass} is obtained. This means that the wave \eref{psis} recovers the classical equations of motion. Thus, the change of ordering \eref{orditr} has no effect over the connection with classical world, for any $r\geq0$ and, in particular, for the non-trivial ordering  \eref{ordzan}.


\section{New bouncing and cyclic solutions}\label{new}
Once the classical equations are obtained, we can now look for solutions that manifest true quantum effects. In this section, we explore two solutions of the new Wheeler-DeWitt equation \eref{eqwdwz}, for $r=1/2$.

\subsection{Bouncing universe I}
\begin{figure}[ttt]
	\centering
	\includegraphics[width=8cm]{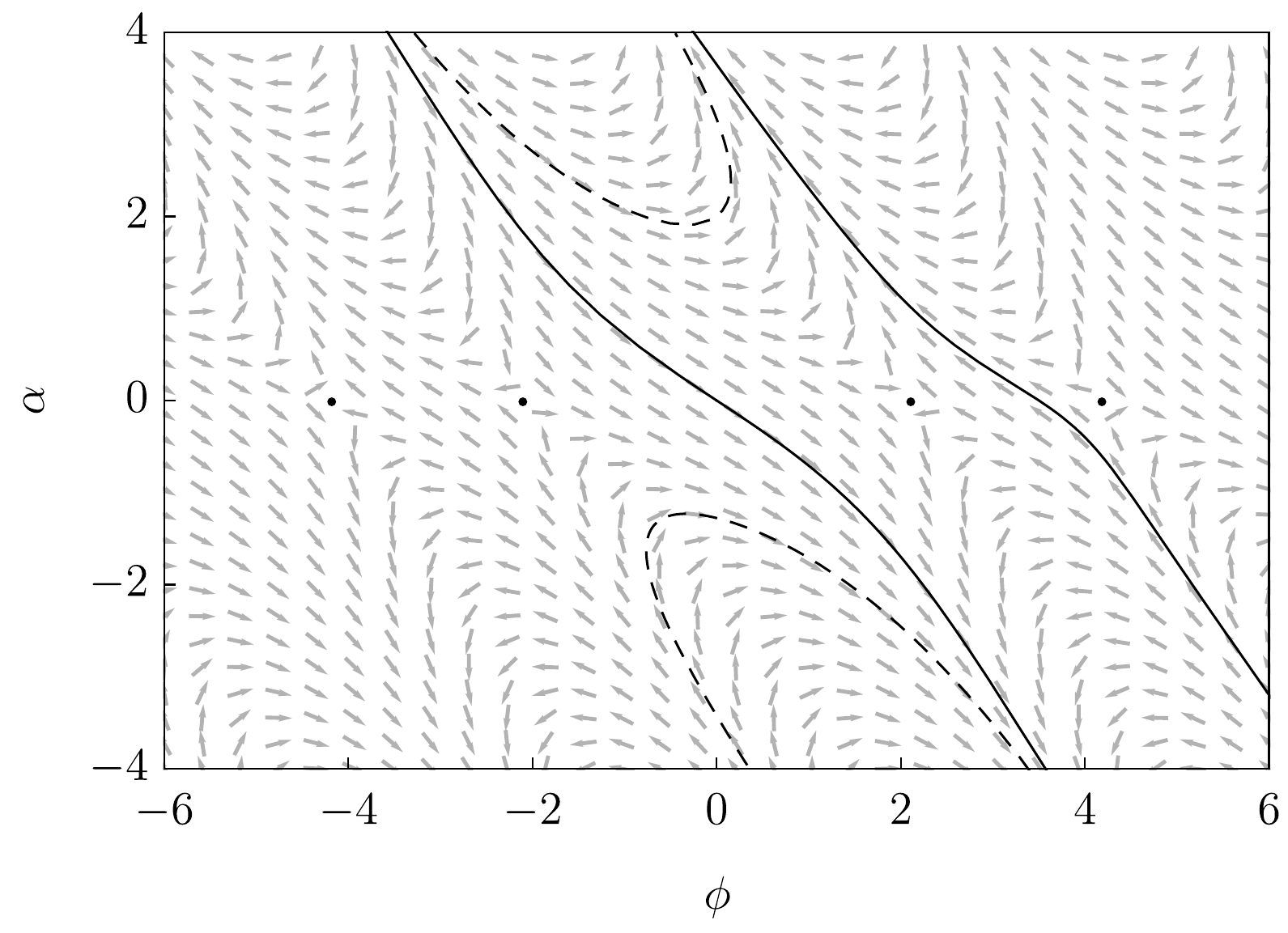}\caption{Phase portrait of the dynamical system  \eref{distdinonger} for the bounce solution  \eref{psibi} with $k=1$. The orientation of the flow represents the time evolution. The dots represent critical points, all of which are saddle points over the straight line $\alpha=0$, given by \eref{bipc}.}
	\label{b1}
\end{figure}

Considering the nontrivial solution of \eref{eqwdwz},
\begin{equation}
\psi_{{\rm BI}}=1+{\rm e}^{3r\alpha}+{\rm e}^{({\rm i}\omega+3r/2)\alpha+{\rm i}k\phi }\;,		\label{psibi}
\end{equation}
a linear combination of \eref{keq0} and \eref{mmk2}, the quantum dynamical system \eref{distdinonger} becomes, for the non-trivial ordering ($r=1/2$):
\begin{equation}\label{bipp}
\eqalign{	\dot{\alpha}=\frac{{\rm e}^{-9\alpha/4}\left[3({\rm e}^{3\alpha/2}-1)\sin\theta-4\omega {\rm e}^{3\alpha/4}+4\omega(1+{\rm e}^{3\alpha/2})\cos\theta\right]}{4\left[ 1+3{\rm e}^{3\alpha/2}+{\rm e}^{3\alpha}+2{\rm e}^{3\alpha/4}(1+{\rm e}^{3\alpha/2})\cos\theta \right]}\;, \cr
	\dot{\phi }=\frac{k{\rm e}^{-9\alpha/4}\left[{\rm e}^{3\alpha/4}+(1+{\rm e}^{3\alpha/2})\cos\theta \right]}{1+3{\rm e}^{3\alpha/2}+{\rm e}^{3\alpha}+2{\rm e}^{3\alpha/4}(1+{\rm e}^{3\alpha/2})\cos\theta}\;,}
\end{equation}
where $\theta\equiv k\phi +\omega\alpha$. The critical points of \eref{bipp} are:
\begin{equation}\label{bipc}
\eqalign{\alpha_{\rm C}=0\;,\cr
	\phi _{\rm C}=\frac{2\pi}{3k}(3n\pm1)\;,}
\end{equation}
where $n\in\mathbb{Z}$.

Figure \ref{b1} shows the phase portrait of dynamical system \eref{bipp}, where we illustrate the possible Bohmian trajectories. The upper dashed curve, with initial conditions $\alpha(0)=2$ and $\phi(0)=0$, represents a bounce because the universe avoids the initial singularity at $a=0$ (equivalent to $\alpha=-\infty$). The dashed curve below, for which $\alpha(0)=-1.5$ and $\phi(0)=-0.75$ represents a universe that expands from the singularity, reaches a maximum, and then contracts back to the singularity (a ``big crunch''). The thick curve on the right represents a singular expanding universe with $\alpha(0)=0$ and $\phi(0)=3.5$. Finally, the thick curve on the left, for which $\alpha(0)=0$ and $\phi(0)=0$, represents a singular contracting universe. Note that the trajectory contains all the the informations about the system and not the particular point where we choose $t=0$, because time, viewed as the parameter of a curve in phase space $\phi\times\alpha$, can always be trivially redefined. From figure \ref{b1}, we can also see that the bounce can only happen if the initial condition $\alpha(t=0)$ is positive, even though that condition is not sufficient. In fact, there are expanding solutions with values $\alpha<0$, but they are all singular.

\begin{figure}[ttt]
	\centering
	\includegraphics[width=8cm]{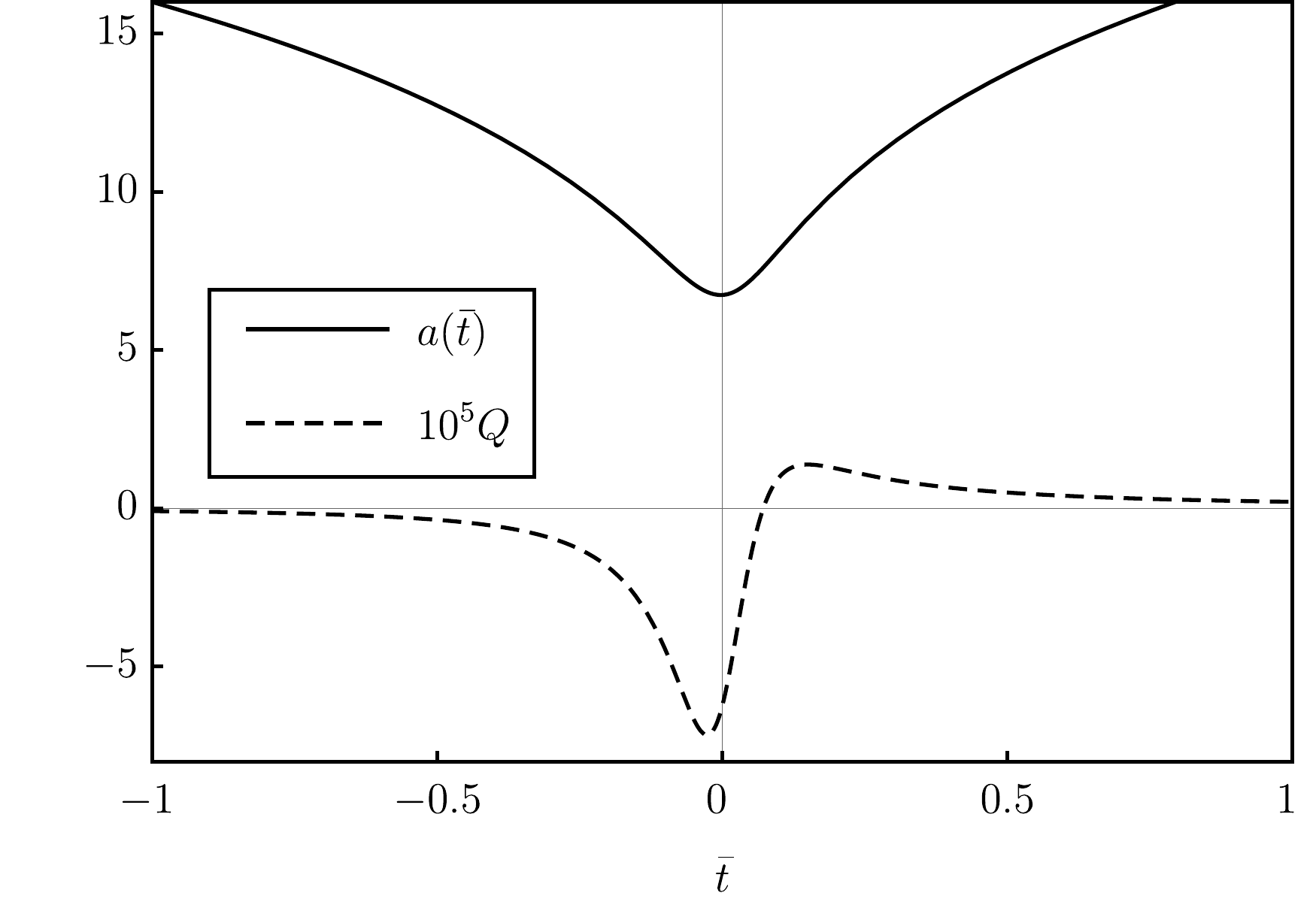}\caption{Qualitative comparison between the time evolution of the scale factor $a={\rm e}^{\alpha}$ and the quantum potential \eref{pqz} for the upper dashed curve in figure \ref{b1}, that represents a bouncing universe. The quantum potential $Q$ was evaluated along that trajectory. For simplicity, time was rescaled by $\bar{t}=(t+900)/15000$ and the global factors on $Q$ were set to unity.}
	\label{b1c1}
\end{figure}

Now we can compare the result above with its analogous for the trivial ordering  \eref{do}. For $r=0$, the dynamical system \eref{distdinonger} for the wave function \eref{psibi} is just
\numparts
\begin{eqnarray}
		\dot{\alpha}&=-k{\rm e}^{-3\alpha}\frac{1+2\cos[k(\alpha+\phi )]}{5+4\cos[k(\alpha+\phi )]}\;,\label{sdqozbiado} \\
		\dot{\phi }&=k{\rm e}^{-3\alpha}\frac{1+2\cos[k(\alpha+\phi )]}{5+4\cos[k(\alpha+\phi )]}\;,\label{sdqozbifido}
\end{eqnarray}
\endnumparts 
since the dispersion relation \eref{omega} degenerates to $\omega=k$ and the real exponential $\exp(3r\alpha)$ in \eref{psibi} becomes a constant. Dividing \eref{sdqozbiado} by \eref{sdqozbifido}, it follows that ${\rm d}\alpha/{\rm d}\phi =-1$. Thus, all solutions are straight lines in the phase space $\phi \times\alpha$ with inclination $-1$. Hence, the system \eref{sdqozbiado} and \eref{sdqozbifido} gives only singular solutions for $a(t)$, no matter what the initial conditions are. Therefore, the wave function \eref{psibi} gives only singular solutions for the standard ordering $r=0$, but it gives singular and bouncing solutions for the non-trivial ordering $r=1/2$. In other words, for the non-trivial factor ordering, it is possible to obtain bounces, which would degenerate to singular solutions for $a$, for the trivial ordering.

To illustrate those new bounce solutions and to compare them with the role played by quantum potential, see figure \ref{b1c1}, where we show one of the non-singular solutions for the non-trivial ordering. It is clear that the modified quantum potential  \eref{pqz} is dominant around $\bar{t}=0$ and smoothly decreases as the universe expands, returning to a classical regime. This shows that the new bounce occurs precisely when its correspondent $Q$ dominates, thus showing its consistency with the Bohmian formalism.

Lastly, the bounce solution found above is stable, in the sense that if we introduced new parameters $c_i\neq0$ to generalize $\psi_{{\rm BI}}$, thus obtaining
\begin{equation}\label{psibitil}
\tilde{\psi}_{{\rm BI}}=c_1+c_2{\rm e}^{3\alpha/2}+c_3{\rm e}^{(\pm{\rm i}\omega+3r/2)\alpha\pm {\rm i}k\phi }\;,
\end{equation}
then the general features illustrated in figure \ref{b1} concerning the possible trajectories would still hold, even for the four sign combinations in the imaginary phase of \eref{psibitil}. That is also true if a $k\neq1$ is chosen, provided that $|k|>3/4$, a limitation imposed by \eref{omega}. All these features can be verified by repeating the process of deriving the dynamical system for $\alpha$ and $\phi$ from the guidance equations  \eref{distdinonger}, but now for $\tilde{\psi}_{{\rm BI}}$.

\subsection{Cyclic universe I}
\begin{figure}[ttt]
	\centering
	\includegraphics[width=9cm]{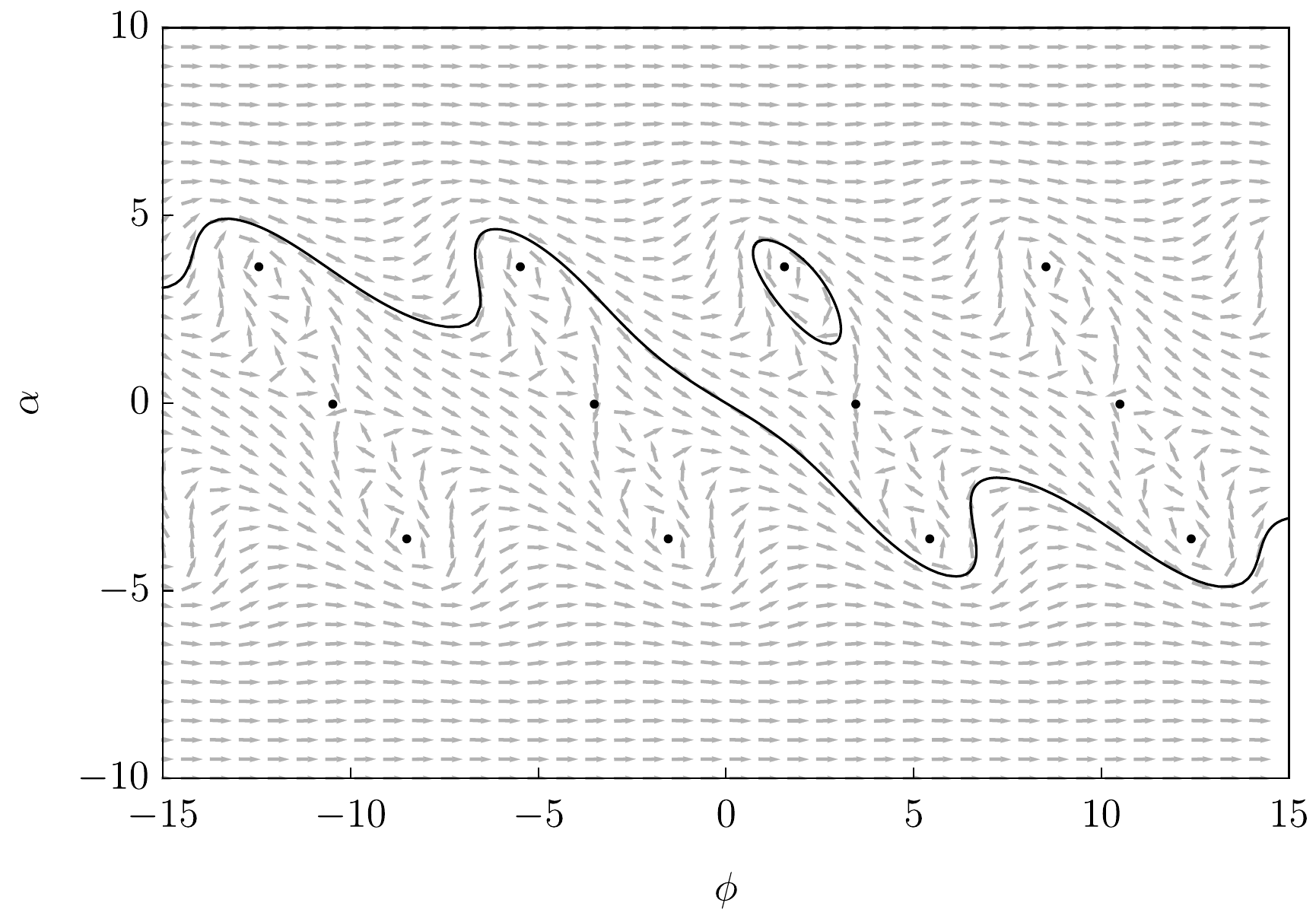}\caption{Phase portrait of the cyclic solution I, obtained from \eref{psici}, for $k=1$ and $k'=0.1$. The dots over the line $\alpha=0$ represent saddle points, given by \eref{cipci}, and the others are center points, given by \eref{cipcii}. All trajectories represent cyclic universes.}
	\label{c1}
\end{figure}
Let us now consider 
\begin{equation}
\psi_{{\rm CI}}={\rm e}^{({\rm i}\omega+3/4)\alpha+{\rm i}k\phi }+{\rm e}^{{\rm i}k'\phi +3\alpha/4}\cosh(\omega'\alpha)\;,\label{psici}
\end{equation}
which is a combination of \eref{mmk2} and  \eref{akl}. The guidance equations \eref{distdinonger} become:
\begin{equation}\label{dsc1}
\eqalign{\dot{\alpha}=-{\rm e}^{-3\alpha}\frac{\omega+\omega\cosh(\omega'\alpha)\cos\beta-\omega'\sinh(\omega'\alpha)\sin\beta}{1+ 2\cosh(\omega'\alpha)\cos\beta+\cosh^2(\omega'\alpha)}\;, \cr
	\dot{\phi }={\rm e}^{-3\alpha}\frac{k+k'\cosh^2(\omega'\alpha)+(k+k')\cosh(\omega'\alpha)\cos\beta}{1+ 2\cosh(\omega'\alpha)\cos\beta+\cosh^2(\omega'\alpha)}\;,}
\end{equation}
where $\beta\equiv(k-k')\phi +\omega\alpha$. There are three classes of critical points. The first one is:
\begin{equation}\label{cipci}
\eqalign{\alpha_{\rm C}=0\;,\cr
	\phi _{\rm C}=\frac{(2n+1)\pi}{k-k'}\;,}
\end{equation}
where $n\in\mathbb{Z}$. The other two are $(\alpha_{C}^{-},\phi _{C}^{+})$ and $(\alpha_{C}^{+},\phi _{C}^{-})$, where
\begin{equation}\label{cipcii}
\eqalign{\alpha_{\rm C}^{\pm}=\pm\frac{1}{\omega'}\cosh^{-1}(\sqrt{y})\;, \cr
	\phi _{\rm C}^{\pm}=\pm\frac{\omega}{k-k'}\alpha_{\rm C}\pm\cos^{-1}\left[ -\frac{k+k'y}{(k+k')\sqrt{y}} \right]+2m\pi\;,}
\end{equation}
where $m\in\mathbb{Z}$ and 
\begin{equation}\label{yy}
y=\frac{1+\frac{k^2}{k'^2}+\frac{\omega^2}{\omega'^2}\pm\sqrt{\left( 1+\frac{k^2}{k'^2}+\frac{\omega^2}{\omega'^2} \right)^2-4\frac{k^2}{k'^2}\left( 1+\frac{\omega^2}{\omega'^2} \right)}}{2(1+\omega^2/\omega'^2)}\;.
\end{equation}
By virtue of \eref{yy}, $k$ and $k'$ must be chosen so that $y$ is real, which is the case for $k=1$ and $k'=0.1$, for instance.

\begin{figure}[ttt]
	\centering
	\includegraphics[width=8cm]{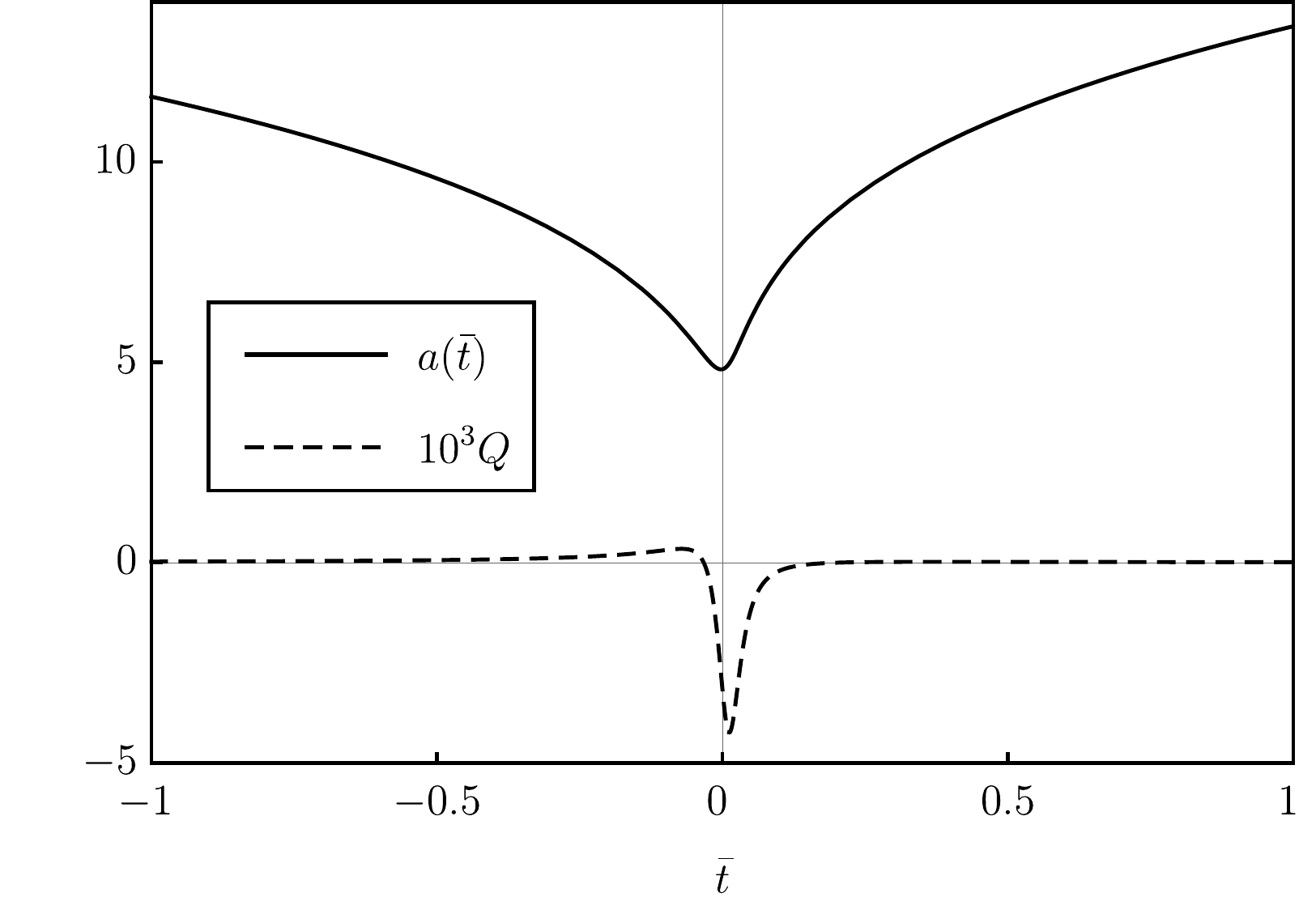}\caption{Qualitative comparison between the scale factor $a=e^{\alpha}$ and the quantum potential  \eref{pqz} for the bottom part of upper right cyclic trajectory of figure \ref{c1}, whose initial conditions are $\alpha(t=0)=\phi(t=0)=2$. The quantum potential is evaluated along that trajectory. For simplicity, time was rescaled by $\bar{t}=(t+210)/2000$ and the global factors on $Q$ were set to unity.}
	\label{c1c1}
\end{figure}

The phase portrait of the dynamical system  \eref{dsc1} is shown in figure \ref{c1}, illustrating two trajectories with cyclic solutions for $\alpha(t)$. For more details about cyclic universes, see \cite{PhysRevD.65.126003}. Those trajectories were obtained by a numerical solution of the dynamical system  \eref{dsc1} for the following initial conditions: $\alpha(0)=\phi (0)=2$ for the upper right cyclic curve and $\alpha(0)=\phi (0)=0$ for the other one. Figure \ref{c1} shows that solution \eref{psici} gives only cyclic universes. The particular behavior depends on the initial conditions. Despite the fact that the solution for which $\alpha(0)=\phi (0)=0$ (and other solutions similar to that one) is not periodic, like the other one, it represents a cyclic universe, since there are alternating phases of expansion and contraction, with a longer contraction phase around $\phi=0$, which is followed by a similar behavior of alternating expansion and contraction phases, as the cyclic character of \ref{c1} suggests.

Observe that for $r=0$ there is no wave function analogous to \eref{psici}, since the condition $|k'|<0=3r/2$ would be impossible, which implies that there is no analogous to \eref{akl} for $r=0$. Hence, the cyclic solutions in figure \ref{c1} are quantum effects only made possible because of the ordering  \eref{orditr}. This remains true as long as $r>0$ and, in particular, for the non-trivial ordering $r=1/2$.

As it was done for $\psi_{{\rm BI}}$, we can see precisely when the quantum effect occurs by comparing the dynamics of the scale factor and the quantum potential for a particular trajectory. In fact, for the cyclic solution on top of figure \ref{c1}, numerical solutions for $a(t)$ and $Q$ give figure \ref{c1c1}, where we can see that the quantum potential is nontrivial precisely when the scale factor bounces from contraction to expansion. That behavior is cyclic and eternal.


\section{Modifications of old solutions}\label{old}
After having studied the new solutions that come from choosing the non-trivial ordering, in this section we address two solutions of the modified Wheler-DeWitt equation \eref{eqwdwz}, for $r=1/2$, that are very similar to their analogues for the original equation \eref{wdwdo}, for the trivial ordering $r=0$. We study the dynamics for both orderings and compare them for two possibilities: bounces and cycles.

\subsection{Bouncing universe II}
\begin{figure}[ttt]
	\centering
	\includegraphics[width=8cm]{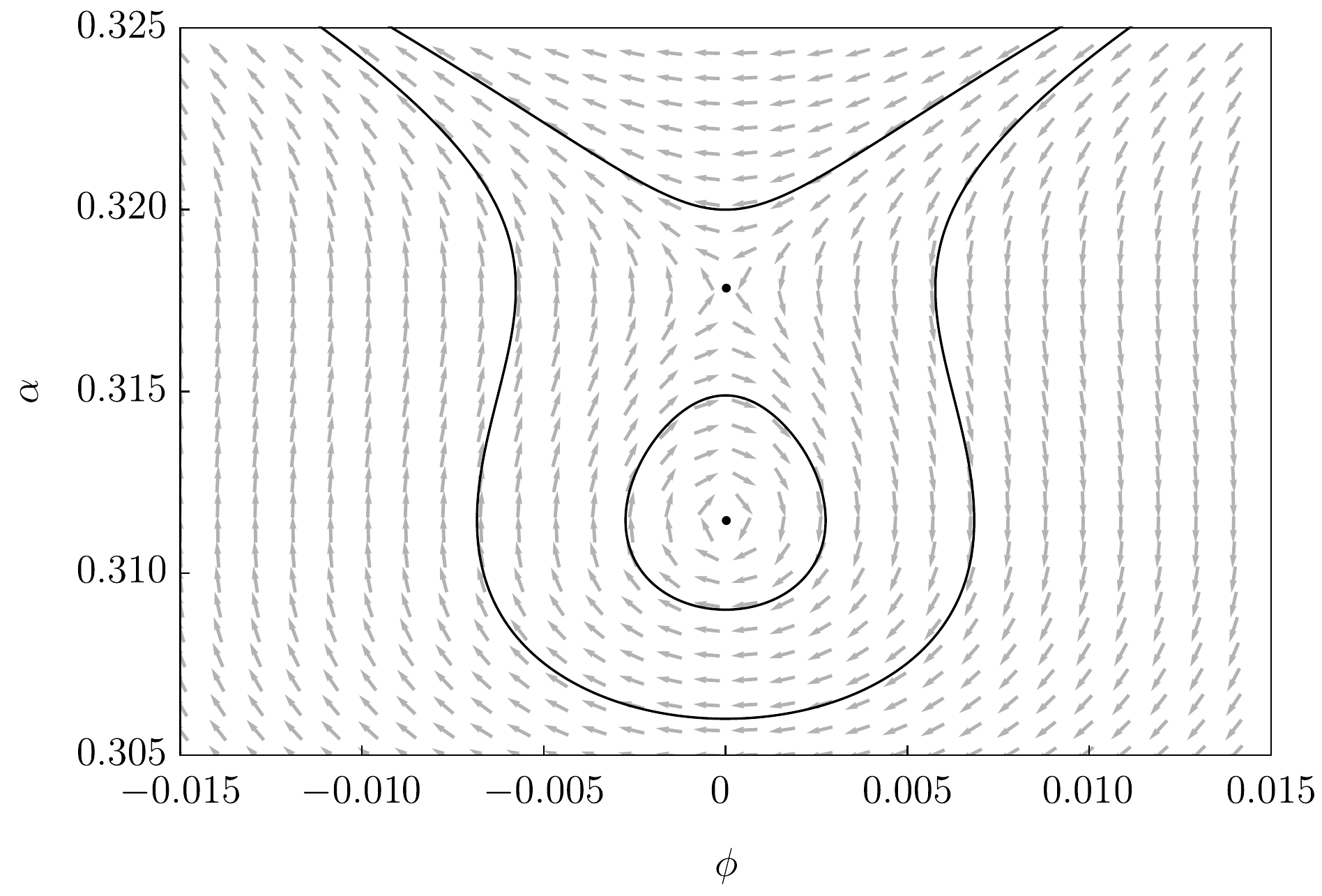}\caption{Phase portrait of the dynamical system  \eref{distdinonger} for the bounce solution  \eref{psibii}, obtained from the non-trivial ordering ($r=1/2$), with $k_0=-5$ and $\sigma=1$. The two dots represent critical points: a center (below) and a saddle point (above).}
	\label{b2}
\end{figure}

The natural generalization of \eref{wdwpo} is the Gaussian wave packet, where $k_0$ and $\sigma$ are real constants:
\begin{equation}
\psi_{{\rm BII}}=\int {\rm d}k\: {\rm e}^{-\frac{(k-k_0)^2}{\sigma^2}}\left[{\rm e}^{({\rm i}\omega+3r/2)\alpha+{\rm i}k\phi }+{\rm e}^{(-{\rm i}\omega+3r/2)\alpha+{\rm i}k\phi }\right]\;, \label{psibii}
\end{equation}
which is a solution of the new Wheeler-DeWitt equation  \eref{eqwdwz}. Because of the new dispersion relation \eref{omega}, we evaluate \eref{psibii} numerically. In \cite{PINTONETO2000194, Colistete:2000ix}, the authors use $k_0=-1$ and $\sigma=1$. But, in our case, since $\omega=[k^2-(3/4)^2]^{1/2}$, which demands $|k|>3/4$, an integration over all real line (like in the standard case) would end up with a solution that mixes two different types of wave solutions, \eref{mmk2} and \eref{akl}; thus, it would not be a rightful generalization of \eref{wdwpo}. We can avoid this problem by assigning another value to $k_0$, the center of the Gaussian weight, so that the effective integration interval lies inside the region $|k|>3/4$.

\begin{figure}[ttt]
	\centering
	\includegraphics[width=8cm]{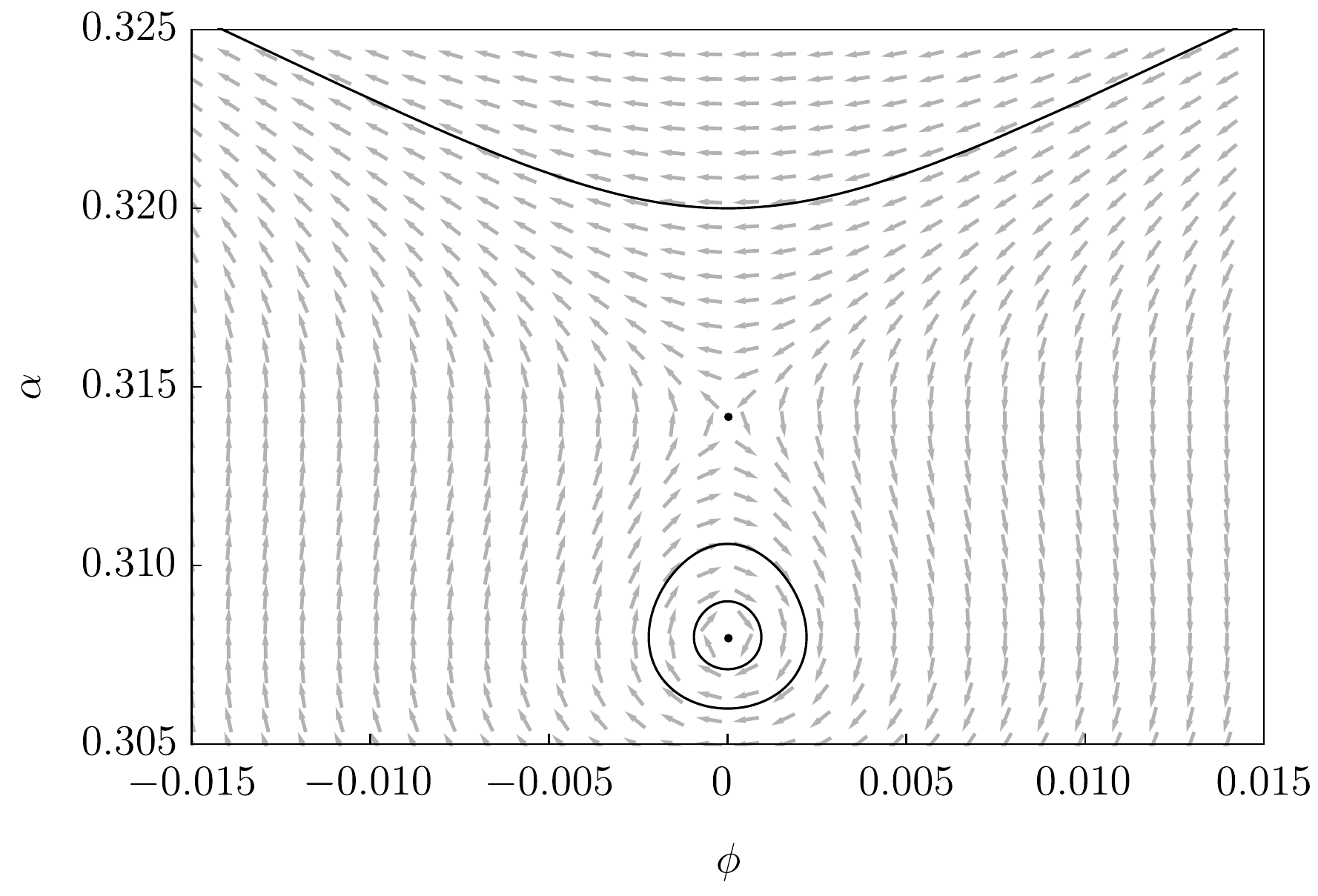}\caption{Phase portrait of the wave packet  \eref{wdwpo}, solution for the trivial ordering $r=0$ (see\cite{PhysRevD.57.4707, PINTONETO2000194, Colistete:2000ix, PINTO-NETO2000}), with $k_0=-5$ and $\sigma=1$. The two dots represent critical points: a center (below) and a saddle point (above).}
	\label{b2n}
\end{figure}

In qualitative terms, the bounce \cite{PINTONETO2000194, Colistete:2000ix} (for \eref{wdwpo}) maintains its physical structure for a different $k_0$ (see figure \ref{b2n} and compare with \cite{PhysRevD.57.4707, PINTONETO2000194, Colistete:2000ix, PINTO-NETO2000}), thus we can evaluate \eref{wdwpo} and \eref{psibii} for say $k_0=-5$, and then compare their respective dynamics for $\phi\times\alpha$. After all these considerations, we can numerically calculate the guidance equations  \eref{distdinonger}, thus obtaining figure \ref{b2}. The correspondent phase portrait of \eref{wdwpo} is given in figure \ref{b2n}. Now, comparing figures \ref{b2} and \ref{b2n}, we can see that the previous result of \cite{PINTONETO2000194, Colistete:2000ix} is qualitatively recovered, in the sense that both dynamical systems have a very similar structure.

In both figures \ref{b2} and \ref{b2n} we show three examples of trajectories obtained from the same three initial conditions. In all initial conditions and critical points, $\phi=0$. For $\alpha(0)=0.306$, the solution is a bounce for the non-trivial ordering (bottom of figure \ref{b2}) and a cycle for trivial ordering (the larger cycle in figure \ref{b2n}). For $\alpha(0)=0.309$, the trajectories are cycles in both orderings: the only cycle in figure \ref{b2} for the non-trivial ordering and the smallest cycle in figure \ref{b2n} for the trivial ordering. Finally, for $\alpha(0)=0.32$, the trajectories are bounces for both orderings. They are shown on the top of figures \ref{b2} and \ref{b2n}. For the region of phase space shown in figure \ref{b2} the critical points have the following approximate coordinates, evaluated numerically: $\phi_{C}\simeq0$ for both; $\alpha_{C}\simeq0.317830$, for the saddle point, and $\alpha_{C}\simeq0.31145$, for the center. For figure \ref{b2n}, the coordinates of the critical points $(0,\alpha_{C})$ are, according to \cite{PINTO-NETO2000}, divided in two groups: for the saddle points, $\alpha_{C}=\pi(2n+1)/2k_0$; for centres, the $\alpha_{C}$'s are the solutions of the transcendental equation $\sigma^2\alpha_{C}=2k_0\cot(k_0\alpha)$.

As a last comment about $\psi_{{\rm BII}}$, we can explain, by simple approximation arguments, why the change of ordering from $r=0$ to $r=1/2$ does not really modifies the dynamics, as it was shown above. In fact, the Gaussian kernel ${\rm exp}[-(k-k_0)^2/\sigma^2]$ gives more weight for the values of $k$ near $k_0$, so we can approximate $\omega(k)$ by (keeping a general ordering parameter $r$)
\begin{equation}
\omega(k)\simeq\omega_0+\omega_1(k-k_0)-\textstyle\frac{1}{2}\omega_2(k-k_0)^2\;,
\end{equation}
where $\omega_0=[k_{0}^{2}-(3r/2)^2]^{1/2}$, $\omega_1=k_0[k_{0}^{2}-(3r/2)^2]^{-1/2}$, and $\omega_2=(3r/2)^2[k_{0}^{2}-(3r/2)^2]^{-3/2}$. For $r=0$ and $k_0=-1$, the original solution  \eref{wdwpo} of \cite{PhysRevD.57.4707,PINTONETO2000194,Colistete:2000ix} is recovered exactly. For $r=1/2$ and $k_0=-5$, we have $\omega_2\sim10^{-3}$, so that $\omega_2(k-k_0)^2$ is negligible. Hence, after the rescaling $\bar{\alpha}\equiv\omega_1\alpha$, the wave packet integral  \eref{psibii} approaches \eref{wdwpo}, up to imaginary phase factors and a global factor of ${\rm exp}[3\alpha/4]$, which is canceled out in the evaluation of $\partial S/\partial q$, using \eref{r-e-s}. Thus, roughly speaking, we can say that the primary effect of the change or ordering in $\psi_{{\rm BII}}$ is a rescaling of $\alpha$, in accordance with what figures \ref{b2} and \ref{b2n} suggest.

\subsection{Cyclic universe II}
Finally, for
\begin{equation}
\psi_{{\rm CII}}=1+{\rm e}^{3r\alpha}+2{\rm e}^{{\rm i}k\phi +3r\alpha/2}\cos(\omega\alpha)\;,\label{psicii}
\end{equation}
which is a combination of \eref{keq0} and \eref{mmk2}, the dynamical system \eref{distdinonger} is, for $r=1/2$:
\begin{equation}\label{dsc2}
\eqalign{\dot{\alpha}=\frac{\textstyle\frac{1}{2}{\rm e}^{-9\alpha/4}\sin(k\phi )\left[ 3({\rm e}^{3\alpha/2}-1)\cos(\omega\alpha)+4\omega({\rm e}^{3\alpha/2}+1)\sin(\omega\alpha) \right]}{1+{\rm e}^{3\alpha}+2{\rm e}^{3\alpha/2}\left[ 2+\cos(2\omega\alpha) \right]+4(1+{\rm e}^{3\alpha/2})\cos(k\phi )\cos(\omega\alpha)}\;, \cr
	\dot{\phi }=\frac{2k{\rm e}^{-9\alpha/4}\cos(\omega\alpha) \left[ ({\rm e}^{3\alpha/2}+1)\cos(k\phi )+2{\rm e}^{3\alpha/4}\cos(\omega\alpha) \right] }{1+{\rm e}^{3\alpha}+2{\rm e}^{3\alpha/2}\left[ 2+\cos(2\omega\alpha) \right]+4(1+{\rm e}^{3\alpha/2})\cos(k\phi )\cos(\omega\alpha)}\;.}
\end{equation}


The critical points of that dynamical system are divided in three sets. The first one is the lattice
\begin{equation}\label{pccii1}
\eqalign{\alpha_{\rm C}=(2n+1)\frac{\pi}{2\omega}\;,\cr
	\phi _{\rm C}=\frac{m\pi}{k}\;,}
\end{equation}
where $m,n\in\mathbb{Z}$. The second is
\begin{equation}\label{pccii2}
\eqalign{\alpha_{\rm C}=0\;,\cr
	\phi _{\rm C}=(2m+1)\frac{\pi}{k}\;,}
\end{equation}
for $m\in\mathbb{Z}$. The third is the set of points $(\phi _{\rm C},\alpha_{\rm C})$ such that $\alpha_{{\rm C}}$ is the solution of the transcendental equation: 
\begin{equation}\label{accii}
\omega\tan(\omega\alpha_{\rm C})+\textstyle\frac{3}{4}\tanh\left( \textstyle\frac{3}{4} \alpha_{\rm C} \right)=0\;,
\end{equation}
and 
\begin{equation}
\phi_{\rm C}=\pm\frac{1}{k}\cos^{-1}\left[ -\frac{\cos(\omega\alpha_{\rm C})}{\cosh(3\alpha_{\rm C}/4)} \right]+\frac{2n\pi}{k}\;.
\end{equation}
Since there are infinitely many $\alpha_{C}$'s satisfying \eref{accii}, the third set of critical points is also a lattice in phase space, but with a varying distance between horizontal sequences of points.

\begin{figure}[ttt]
	\centering
	\includegraphics[width=8cm]{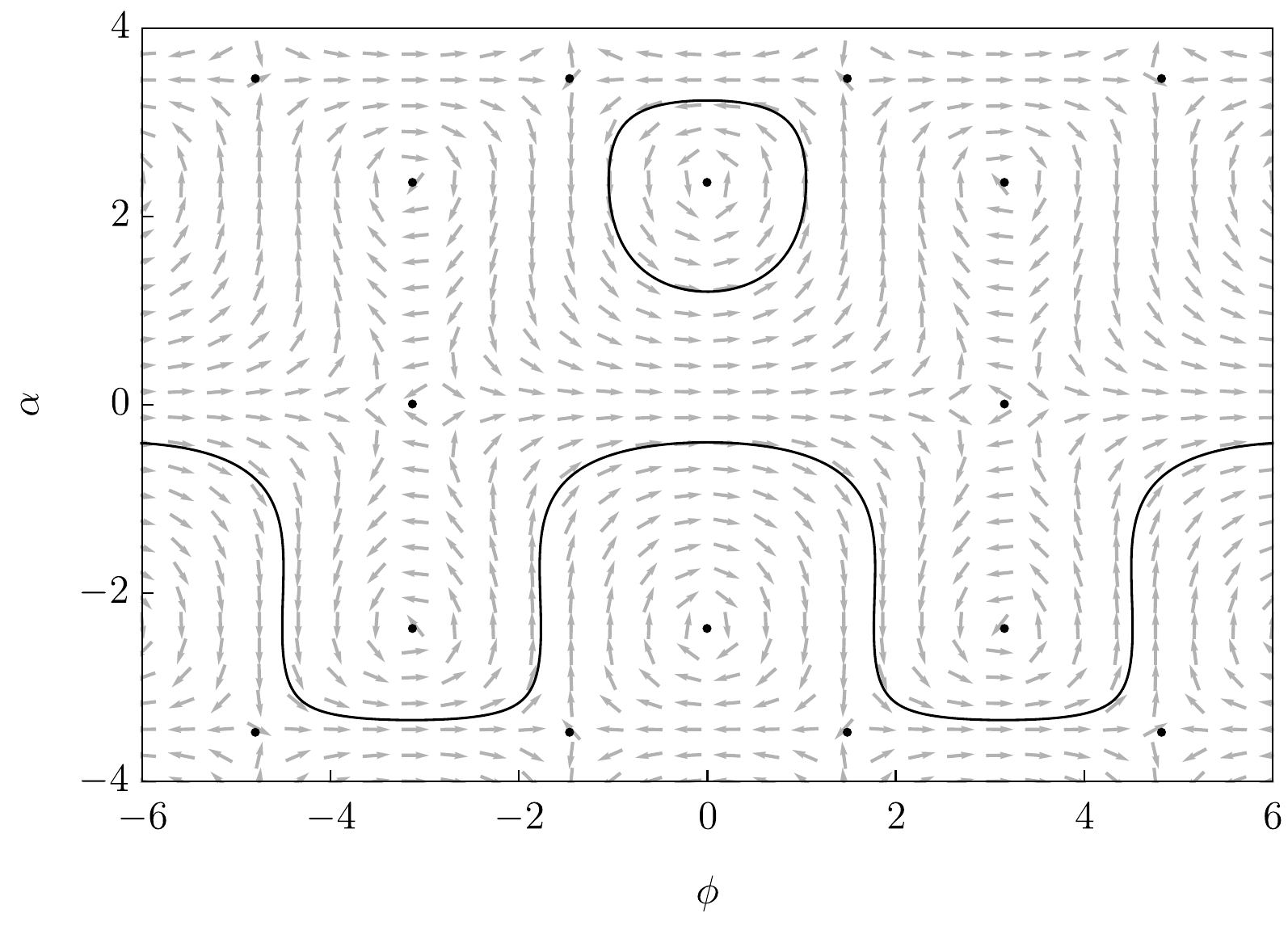}\caption{Phase portrait of the cyclic solution II, \eref{psicii}, for $k=1$. The dots represent critical points: the center points are given by \eref{pccii1} and the others are saddle points. All trajectories represent cyclic universes.}
	\label{c2}
\end{figure}

The phase portrait of the dynamical system  \eref{dsc2} is shown in figure \ref{c2}, that illustrates two trajectories with a cyclic solution for $\alpha(t)$. These trajectories were numerically obtained from dynamical system \eref{dsc2}, for the following initial conditions: $\alpha(0)=1.2$ and $\phi (0)=0$ for the upper cyclic curve and $\alpha(0)=-0.4$ and $\phi (0)=0$ for the other curve.

The above solutions would be very similar for the standard ordering  \eref{do}. In fact, the quantum dynamical system  \eref{distdinonger} becomes, for wave solution \eref{psicii} with $r=0$,
\begin{equation}\label{dsc2do}
\eqalign{\dot{\alpha}=\frac{2k{\rm e}^{-3\alpha}\sin(k\alpha)\sin(k\phi )}{3+\cos(2k\alpha)+4\cos(k\alpha)\cos(k\phi )}\;,\cr
	\dot{\phi }=\frac{2k{\rm e}^{-3\alpha}\cos(k\alpha)\left[\cos(k\alpha)+\cos(k\phi )\right]}{3+\cos(2k\alpha)+4\cos(k\alpha)\cos(k\phi )}\;.}
\end{equation}


\begin{figure}[ttt]
	\centering
	\includegraphics[width=8cm]{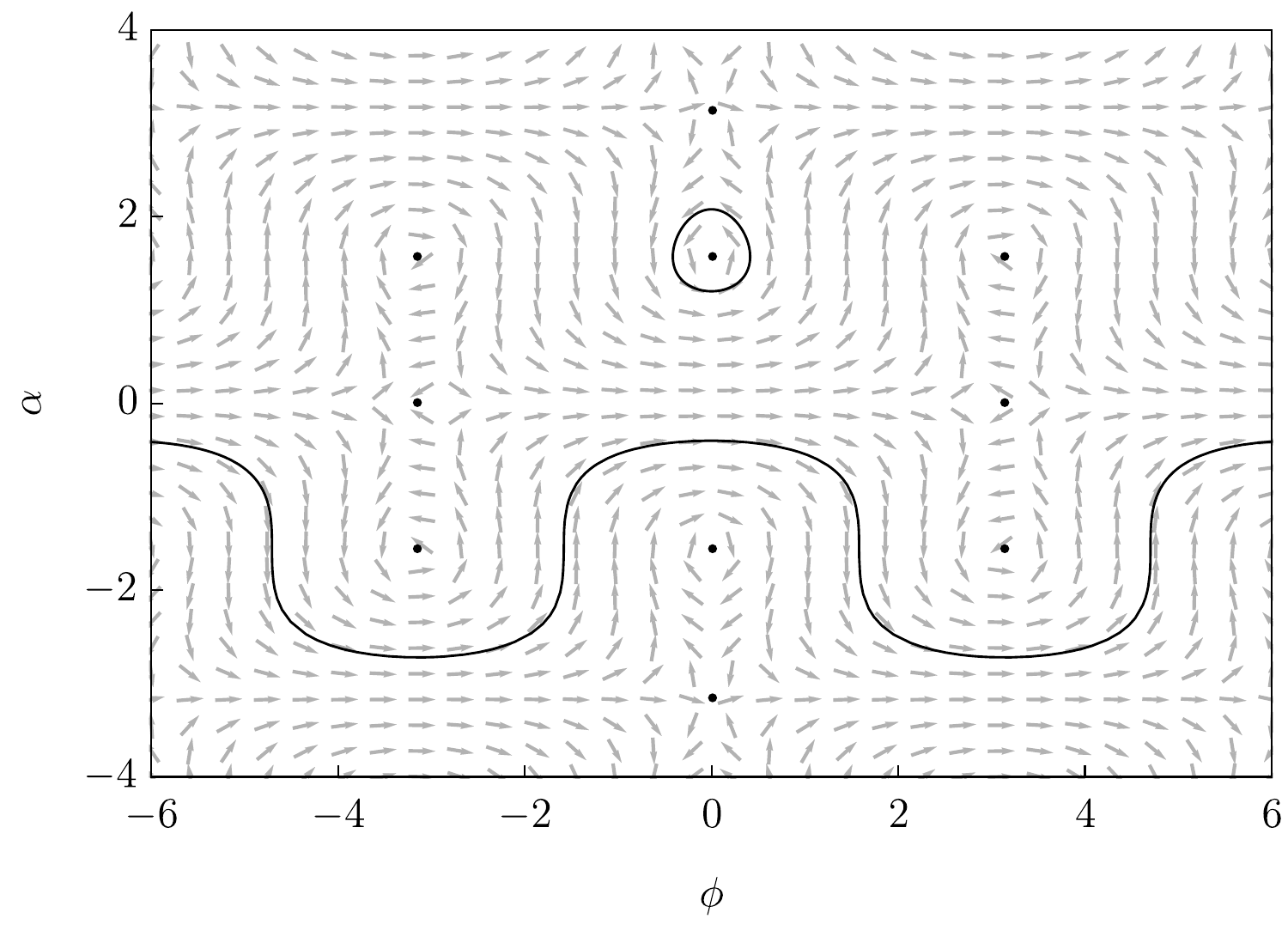}\caption{Phase portrait of the cyclic solution II, obtained from \eref{psicii}, for $k=1$, with the trivial ordering $r=0$. The dots represent critical points: \eref{c2nps1} and  \eref{c2nps2} are saddle points and \eref{c2npc} are centers. All trajectories represent cyclic universes.}
	\label{c2n}
\end{figure}
The critical points of  \eref{dsc2do} are easy to find:
\numparts\label{dsc2dopc}
\begin{eqnarray}
	\phi_{C}&=\frac{2\pi m}{k}\;,	&\qquad\alpha_{C}=(2n-1)\frac{\pi}{k}\;;\label{c2nps1}\\
	\phi_{C}&=(2m-1)\frac{\pi}{k}\;,&\qquad\alpha_{C}=\frac{2\pi n}{k}\;;\label{c2nps2} \\
	\phi_{C}&=\frac{\pi m}{k}\;, &\qquad\alpha_{C}=\left(n+\frac{1}{2}\right)\frac{\pi}{k}\;; \label{c2npc}
\end{eqnarray}
\endnumparts
where $m,n\in\mathbb{Z}$. The phase portrait of dynamical system \eref{dsc2do} is given in figure \ref{c2n}, where we show two trajectories with the same initial conditions as the ones in figure \ref{c2}. It then becomes clear that \eref{psicii} gives very similar dynamics for the orderings considered here. 


\section{Conclusions}\label{conc}
In summary, we can say that the general structure of the Bohmian approach to the quantum cosmology of  \eref{l} indeed stills valid for the class of orderings  \eref{orditr}, in agreement with the argument of \cite{PINTO-NETO2000}, but the description of the expansion of the universe itself, for given initial conditions $\alpha(t=0),\phi(t=0)$, changes. More precisely, the modified Hamilton-Jacobi structure of \eref{wdwdore} is valid for all orderings of the form \eref{orditr}, but both the wave functions and the quantum potential actually depend on the choice of factor ordering, as we have shown above. The connection between classical and quantum dynamics is also maintained, for two reasons. First, the trivial plane wave solution  \eref{psis}, for which $R=\sqrt{\psi^{*}\psi}$ is constant (which implies $Q=0$), leads to the classical equations of motion \eref{dinclass}, for any $r\geq0$. Second, even if a non-trivial wave function $\psi$ is considered, if follows from \eref{pqz} that $Q\sim\hbar^2a^{-3}$ (for any $r\geq0$), so that in an expanding universe, \eref{wdwdore} degenerates to classical Hamilton-Jacobi equation, since $Q\rightarrow0$.  

The differences between the trivial ordering $r=0$ and the non-trivial ordering $r=1/2$ for the theory investigated above are divided in two classes. First, in section \ref{new}, we have shown that new bouncing and cyclic universes become possible by new solutions of Wheeler-DeWitt equation. Those solutions, $\psi_{{\rm BI}}$ and $\psi_{{\rm CI}}$, can be constructed from the modified equation \eref{eqwdwz} if, and only if, $r\neq0$. In other words, those new solutions are impossible for the trivial ordering that leads to \eref{wdwdo}. The second class of solutions are $\psi_{{\rm BII}}$ and $\psi_{{\rm CII}}$, studied in section \ref{old}. Those solutions were already possible for \eref{wdwdo}, and we have shown that the modification of factor ordering only provides a slight modification of them. Therefore, we can say in conclusion that the non-trivial ordering opens new possibilities for quantum cosmology with Bohm-de Broglie interpretation. 

We would like to stress that our results are valid for the background cosmology of \eref{ham}. Since the theory of quantum cosmological perturbations for \eref{ham} was investigated in \cite{PhysRevD.95.023523, PhysRevD.97.083517, PhysRevD.95.123522} for the trivial ordering \eref{do}, it is quite natural to ask what are the consequences of the alternative description developed here for those perturbations. This is an open question for future works.

Finally, a remark about the connection with the usual interpretation of quantum mechanics. As we know, the Bohmian interpretation allows one to deal with non-normalizable functions, which may be problematic in the usual one. Although, for suitable normalizable functions, we can define the inner product
\begin{equation}
	\langle\varphi | \psi\rangle=\int\varphi^{*}\psi \,{\rm e}^{3(1-r)\alpha}{\rm d}\alpha{\rm d}\phi\; ,
\end{equation}
where the measure ${\rm d}\mu_r=\exp[3(1-r)\alpha]{\rm d}\alpha{\rm d}\phi$ ensures that the Hamiltonian operator $\hat{\mathcal{H}}_{r}$, defined by
\begin{equation}
	\hat{\mathcal{H}}_{r}=\textstyle \frac{1}{2}{\rm e}^{-3\alpha}\left( \partial_{\alpha\alpha}-3r\partial_{\alpha}-\partial_{\phi\phi} \right)\;,
\end{equation}
 is Hermitian:
\begin{equation}
		\langle\varphi | \hat{\mathcal{H}}_r | \psi\rangle = \langle\psi | \hat{\mathcal{H}}_r | \varphi \rangle^{*} \;.
\end{equation}
Moreover, returning again to the causal interpretation, the concept of local expectation value of an observable $\hat{A}$, defined by
\begin{equation}
	\langle A\rangle_{\rm{local}}\equiv\rm{Re}\left(\frac{\psi^{*}\hat{A}\psi}{\psi^{*}\psi}\right)\;,
\end{equation}
may replace the expectation value of an observable for a more general class of wave functions (including non-normalizable ones), thus completing the connection with usual interpretation.

\ack
We thank to Hideki Maeda, Ingrid Ferreira da Costa, and Jorge Zanelli for very important discussions about this paper. This study was financed in part by the \emph{Coordena\c{c}\~ao de Aperfei\c{c}oamento de Pessoal de N\'ivel Superior} - Brazil (CAPES) - Finance Code 001 and also by FAPES and CNPq from Brazil. OFP thanks the Alexander von Humboldt foundation for funding and the Institute for Theoretical Physics of the Heidelberg University for kind hospitality.

\section*{References}
\bibliography{biblio}

\providecommand{\newblock}{}
\begin{thebibliography}{10}
\expandafter\ifx\csname url\endcsname\relax
  \def\url#1{{\tt #1}}\fi
\expandafter\ifx\csname urlprefix\endcsname\relax\def\urlprefix{URL }\fi
\providecommand{\eprint}[2][]{\url{#2}}

\bibitem{Riess:1998cb}
Riess A {\em et~al.\/} (Supernova Search Team) 1998 {\em Astron. J.\/} {\bf
  116} 1009--1038 (\textit{Preprint} \eprint{astro-ph/9805201})

\bibitem{Perlmutter:1998np}
Perlmutter S {\em et~al.\/} (Supernova Cosmology Project) 1999 {\em Astrophys.
  J.\/} {\bf 517} 565--586 (\textit{Preprint} \eprint{astro-ph/9812133})

\bibitem{Ratra:1987rm}
{B Ratra, PJE Peebles} 1988 {\em Phys. Rev.\/} {\bf D37} 3406

\bibitem{Caldwell:1997ii}
{RR Caldwell, R Dave, PJ Steinhardt} 1998 {\em Phys. Rev. Lett.\/} {\bf 80}
  1582--1585 (\textit{Preprint} \eprint{astro-ph/9708069})

\bibitem{Faraoni:2004pi}
Faraoni V 2004 {\em {Cosmology in scalar tensor gravity}\/} vol 139 ISBN
  1402019882

\bibitem{Linde1990xn}
Linde A 1990 {\em {Inflation and quantum cosmology. Boston, USA: Academic}\/}
  ISBN 9780124336933

\bibitem{PhysRevD.57.4707}
{R Colistete Jr, JC Fabris, N Pinto-Neto} 1998 {\em Phys. Rev. D\/} {\bf 57}(8)
  4707--4717 (\textit{Preprint} \eprint{gr-qc/9711047})

\bibitem{PINTONETO2000194}
{N Pinto-Neto, AF Velasco, R Colistete} 2000 {\em Physics Letters A\/} {\bf
  277} 194 -- 204 ISSN 0375-9601 (\textit{Preprint} \eprint{gr-qc/0001074})

\bibitem{Colistete:2000ix}
{R Colistete Jr, JC Fabris, N Pinto-Neto} 2000 {\em Phys. Rev. D\/} {\bf 62}
  083507 (\textit{Preprint} \eprint{gr-qc/0005013})

\bibitem{PhysRevD.96.063502}
{S Colin, N Pinto-Neto} 2017 {\em Phys. Rev. D\/} {\bf 96}(6) 063502

\bibitem{Horndeski:1974wa}
\relax GW~Horndeski 1974 {\em Int. J. Theor. Phys.\/} {\bf 10} 363--384

\bibitem{Ostrogradsky:1850fid}
Ostrogradsky M 1850 {\em Mem. Acad. St. Petersbourg\/} {\bf 6} 385--517

\bibitem{Kobayashi_2019}
Kobayashi T 2019 {\em Reports on Progress in Physics\/} {\bf 82} 086901
  (\textit{Preprint} \eprint{1901.07183})

\bibitem{PhysRevLett.119.161101}
{LIGO Scientific Collaboration and Virgo Collaboration (BP Abbott {\it et al})}
  2017 {\em Phys. Rev. Lett.\/} {\bf 119}(16) 161101 (\textit{Preprint}
  \eprint{1710.05832})

\bibitem{Goldstein_2017}
{A Goldstein {\it et al}} 2017 {\em The Astrophysical Journal\/} {\bf 848} L14
  (\textit{Preprint} \eprint{1710.05446})

\bibitem{Abbott_2017}
{B P Abbott {\it et al}} 2017 {\em The Astrophysical Journal\/} {\bf 848} L13
  (\textit{Preprint} \eprint{1710.05834})

\bibitem{Omnes}
Omn\`es R 1994 {\em {The {I}nterpretation of {Q}uantum {M}echanics}\/}
  (Princeton, NJ: Princeton University Press)

\bibitem{ACACIODEBARROS1998229}
{JA de Barros, N Pinto-Neto, MA Sagioro-Leal} 1998 {\em Physics Letters A\/}
  {\bf 241} 229 -- 239 ISSN 0375-9601 (\textit{Preprint}
  \eprint{gr-qc/9710084})

\bibitem{10.2307/193027}
{C Callender, R Weingard} 1994 {\em PSA: Proceedings of the Biennial Meeting of
  the Philosophy of Science Association\/} {\bf 1994} 218--227 ISSN 02708647
  \urlprefix\url{https://www.jstor.org/stable/193027}

\bibitem{PintoNeto:2004uf}
{N Pinto-Neto} 2005 {\em Found. Phys.\/} {\bf 35} 577--603 (\textit{Preprint}
  \eprint{gr-qc/0410117})

\bibitem{Blaut_1996}
{A B{\l}aut, JK Glikman} 1996 {\em Classical and Quantum Gravity\/} {\bf 13}
  39--49

\bibitem{Barros_1997}
{J Acacio de Barros, N Pinto-Neto} 1997 {\em Classical and Quantum Gravity\/}
  {\bf 14} 1993--1995

\bibitem{PINTO-NETO2000}
{N Pinto-Neto} 2000 {\em {Brazilian Journal of Physics}\/} {\bf 30} 330 -- 345
  ISSN 0103-9733

\bibitem{PhysRev.85.166}
Bohm D 1952 {\em Phys. Rev.\/} {\bf 85}(2) 166--179

\bibitem{PhysRev.85.180}
Bohm D 1952 {\em Phys. Rev.\/} {\bf 85}(2) 180--193

\bibitem{Kiefer:2013jqa}
Kiefer C 2013 {\em ISRN Math. Phys.\/} {\bf 2013} 509316 (\textit{Preprint}
  \eprint{1401.3578})

\bibitem{andprobtime}
Anderson E 2012 {\em Annalen der Physik\/} {\bf 524} 757--786

\bibitem{cushing2013bohmian}
{JT Cushing, A Fine, S Goldstein} 2013 {\em Bohmian {Mechanics and Quantum
  Theory: An Appraisal}\/} Boston Studies in the Philosophy and History of
  Science (Springer Netherlands) ISBN 9789401587150

\bibitem{durr2009bohmian}
{D D{\"u}rr, S Teufel} 2009 {\em Bohmian {Mechanics: The Physics and
  Mathematics of Quantum Theory}\/} (Springer Berlin Heidelberg) ISBN
  9783540893448

\bibitem{holland_1993}
\relax PR~Holland 1993 {\em {The Quantum Theory of Motion: An Account of the de
  Broglie-Bohm Causal Interpretation of Quantum Mechanics}\/} (Cambridge
  University Press)

\bibitem{freire2014quantum}
Freire O 2014 {\em The {Quantum Dissidents: Rebuilding the Foundations of
  Quantum Mechanics} (1950-1990)\/} (Springer Berlin Heidelberg) ISBN
  9783662446621

\bibitem{pladevall2019applied}
{XO Pladevall, J Mompart} 2019 {\em Applied {Bohmian Mechanics: From Nanoscale
  Systems to Cosmology}\/} (Jenny Stanford Publishing) ISBN 9781000650105

\bibitem{HOLLAND199395}
Holland P 1993 {\em Physics Reports\/} {\bf 224} 95 -- 150 ISSN 0370-1573

\bibitem{PhysRevD.29.2738}
{T Christodoulakis, J Zanelli} 1984 {\em Phys. Rev. D\/} {\bf 29}(12)
  2738--2745

\bibitem{Christodoulakis1986}
{T Christodoulakis, J Zanelli} 1986 {\em Il Nuovo Cimento B (1971-1996)\/} {\bf
  93} 1--21 ISSN 1826-9877

\bibitem{Maeda_2015}
Maeda H 2015 {\em Classical and Quantum Gravity\/} {\bf 32} 235023

\bibitem{Calcagni:2017sdq}
Calcagni G 2017 {\em {Classical and Quantum Cosmology}\/} Graduate Texts in
  Physics (Springer)

\bibitem{PhysRevD.97.083517}
{AP Bacalhau, N Pinto-Neto, SDP Vitenti} 2018 {\em Phys. Rev. D\/} {\bf 97}(8)
  083517 (\textit{Preprint} \eprint{1706.08830})

\bibitem{doi:10.1142/S0218271898000164}
{JA de Barros, N Pinto-Neto} 1998 {\em International Journal of Modern Physics
  D\/} {\bf 07} 201--213 (\textit{Preprint}
  \eprint{https://doi.org/10.1142/S0218271898000164})
  \urlprefix\url{https://doi.org/10.1142/S0218271898000164}

\bibitem{NOVELLO2008127}
{M Novello, SEP Bergliaffa} 2008 {\em Physics Reports\/} {\bf 463} 127 -- 213
  ISSN 0370-1573 (\textit{Preprint} \eprint{0802.1634})

\bibitem{PhysRevD.95.023523}
{DCF Celani, N Pinto-Neto, SDP Vitenti} 2017 {\em Phys. Rev. D\/} {\bf 95}(2)
  023523 (\textit{Preprint} \eprint{1610.04933})

\bibitem{PhysRevD.95.123522}
{N Pinto-Neto, A Scardua} 2017 {\em Phys. Rev. D\/} {\bf 95}(12) 123522
  (\textit{Preprint} \eprint{1701.07670})

\bibitem{PhysRev.114.1182}
Anderson J 1959 {\em Phys. Rev.\/} {\bf 114}(4) 1182--1184

\bibitem{PhysRevD.20.830}
Komar A 1979 {\em Phys. Rev. D\/} {\bf 20}(4) 830--833

\bibitem{PhysRevD.37.888}
Vilenkin A 1988 {\em Phys. Rev. D\/} {\bf 37}(4) 888--897

\bibitem{PhysRevA.41.1199}
{ST Ali, H-D Doebner} 1990 {\em Phys. Rev. A\/} {\bf 41}(3) 1199--1210

\bibitem{PhysRevD.59.063513}
{N Kontoleon, DL Wiltshire} 1999 {\em Phys. Rev. D\/} {\bf 59}(6) 063513

\bibitem{PhysRevD.80.103507}
{F Amemiya, T Koike} 2009 {\em Phys. Rev. D\/} {\bf 80}(10) 103507

\bibitem{PhysRevD.100.046008}
{T Demaerel, W Struyve} 2019 {\em Phys. Rev. D\/} {\bf 100}(4) 046008

\bibitem{Steigl:2005fk}
{R \u Steigl, F Hinterleitner} 2006 {\em Class. Quant. Grav.\/} {\bf 23}
  3879--3894 (\textit{Preprint} \eprint{gr-qc/0511149})

\bibitem{gotay1999}
Gotay M 1999 {\em Journal of Mathematical Physics\/} {\bf 40} 2107--2116

\bibitem{Chithiika}
{V Chithiika Ruby, VK Chandrasekar, M Senthilvelan, M Lakshmanan} 2015 {\em
  Journal of Mathematical Physics\/} {\bf 56} 012103

\bibitem{goldstein2002classical}
Goldstein H 2002 {\em Classical {M}echanics\/} (Pearson Education) ISBN
  9788177582833

\bibitem{PhysRevD.65.126003}
{PJ Steinhardt, N Turok} 2002 {\em Phys. Rev. D\/} {\bf 65}(12) 126003
  (\textit{Preprint} \eprint{hep-th/0111098})

\end{thebibliography}
\end{document}